\documentclass[twocolumn,tighten]{aastex63}

\hypersetup{linkcolor=red,citecolor=magenta,urlcolor=blue}

\usepackage{amsmath}
\usepackage[noabbrev]{cleveref}
\usepackage{hhline}
\usepackage{makecell}
\usepackage{savesym}
\savesymbol{tablenum}
\usepackage{siunitx}
\restoresymbol{SIX}{tablenum}

\usepackage{tikz}
\usepackage{mathdots}
\usepackage{yhmath}
\usepackage{cancel}
\usepackage{color}
\usepackage{array}
\usepackage{multirow}
\usepackage{amssymb}
\usepackage{gensymb}
\usepackage{tabularx}
\usepackage{booktabs}
\usetikzlibrary{fadings}
\usetikzlibrary{patterns}
\usetikzlibrary{shadows.blur}

\newcommand{\omegacen}{$\omega$\,Centauri}

\submitjournal{ApJ}

\shorttitle{Brown dwarfs in \omegacen{}}
\shortauthors{Gerasimov et al.}

\graphicspath{{./}{figures/}}

\begin{document}

\title{The \textit{HST} large programme on \omegacen{} -- V. Exploring the Ultracool Dwarf Population with Stellar Atmosphere and Evolutionary Modelling}

\correspondingauthor{Roman Gerasimov}
\email{romang@ucsd.edu}

\author[0000-0003-0398-639X]{Roman Gerasimov}
\affiliation{Center for Astrophysics and Space Sciences,
University of California, San Diego, La Jolla, California 92093, USA
}

\author[0000-0002-6523-9536]{Adam J.\ Burgasser}
\affiliation{Center for Astrophysics and Space Sciences,
University of California, San Diego, La Jolla, California 92093, USA
}

\author[0000-0002-8546-9128]{Derek Homeier}
\affiliation{F{\"o}rderkreis Planetarium G{\"o}ttingen, 37085 G{\"o}ttingen, Germany}
\affiliation{Aperio Software Ltd., Insight House, Riverside Business Park, Stoney Common Road,
Stansted, Essex, CM24 8PL, United Kingdom}

\author[0000-0003-4080-6466]{Luigi R.\ Bedin}
\affiliation{INAF—Osservatorio Astronomico di Padova, Vicolo dell’Osservatorio 5, I-35122 Padova, Italy
}

\author[0000-0002-5376-3883]{Jon M.\ Rees}
\affiliation{Lick Observatory, 
P.O.\ Box 85, Mt.\ Hamilton, California 95140, USA
}

\author[0000-0001-8834-3734]{Michele Scalco}
\affiliation{Dipartimento di Fisica e Scienze della Terra, Università di Ferrara, Via Giuseppe Saragat 1, Ferrara I-44122, Italy
}
\affiliation{INAF—Osservatorio Astronomico di Padova, Vicolo dell’Osservatorio 5, I-35122 Padova, Italy
}

\author{Jay Anderson}
\affiliation{Space Telescope Science Institute, 3700 San Martin Drive, Baltimore, MD 21218, USA
}

\author{Maurizio Salaris}
\affiliation{Astrophysics Research Institute, Liverpool John Moores Univ., IC2, Liverpool Science Park, 146 Brownlow Hill, Liverpool, L3 5RF, UK
}
\affiliation{INAF – Osservatorio Astronomico di Abruzzo, Via M. Maggini, s/n, I-64100, Teramo, Italy}

\begin{abstract}
Brown dwarfs can serve as both clocks and chemical tracers of the evolutionary history of the Milky Way due to their continuous cooling and high sensitivity of spectra to composition. We focus on brown dwarfs in globular clusters that host some of the oldest coeval populations in the galaxy. Currently, no brown dwarfs in globular clusters have been confirmed, but they are expected to be uncovered with advanced observational facilities such as \textit{JWST}. In this paper we present a new set of stellar models specifically designed to investigate low-mass stars and brown dwarfs in \omegacen{} -- the largest known globular cluster. The parameters of our models were derived from iterative fits to \textit{HST} photometry of the Main Sequence members of the cluster. Despite the complex distribution of abundances and the presence of multiple Main Sequences in \omegacen{}, we find that the modal colour-magnitude distribution can be represented by a single stellar population with parameters determined in this study. The observed luminosity function is well-represented by two distinct stellar populations having solar and enhanced helium mass fractions and a common initial mass function, in agreement with previous studies. Our analysis confirms that the abundances of individual chemical elements play a key role in determining the physical properties of low-mass cluster members. We use our models to draw predictions of brown dwarf colours and magnitudes in anticipated \textit{JWST} \textit{NIRCam} data, confirming that the beginning of the substellar sequence should be detected in \omegacen{} in forthcoming observations.
\end{abstract}


\section{Introduction} \label{sec:1_introduction}
Over $1/6$ \citep{BD_abundance_2,BD_abundance} of the local stellar population consists of brown dwarfs -- substellar objects with masses below the threshold for sustained hydrogen fusion ($\gtrsim 0.07\ \mathrm{M}_\odot$ for solar composition, \citealt{1963PThPh..30..460H,1962AJ.....67S.579K,1963ApJ...137.1121K,BD_mass_limit}). In contrast to hydrogen-burning stars, brown dwarfs do not establish energy equilibrium and begin cooling continuously shortly after formation, gradually decreasing in effective temperature and luminosity. The characteristically low effective temperatures of such objects ($T_\mathrm{eff} \lesssim 3000\ \mathrm{K}$) allow complex molecular chemistry to take place in their atmospheres, which evolves throughout the cooling process as compounds with lower dissociation energy form. At sufficiently low temperatures, species condense into liquid and solid forms, forming {clouds} of various compositions \citep{first_dust,1996A&A...305L...1T,2002ApJ...568..335M}. The resulting sensitivity of spectra to elemental abundances and age (through cooling) imply that brown dwarfs have the potential to be used as chemical tracers for studies of galactic populations and the Milky Way at large \citep{2009IAUS..258..317B,2020ApJ...892...31B}. Furthermore, the unusual physical conditions characteristic of brown dwarfs, including their low effective temperatures, high densities \citep{BD_density}, and partially degenerate, fully convective interiors 
\citep{MS_inflection_0,1993RvMP...65..301B} provide empirical tests for studies of matter in extreme conditions \citep{metallic_hydrogen,2020NatPh..16..432H}, cloud formation in exoplanetary atmospheres \citep{2014Natur.505...69K,exo_atmospheres}, and even searches for physics beyond the Standard Model \citep{pp_chain_beyond_SM}.

Unfortunately, the faint luminosities and low temperatures of brown dwarfs make these objects challenging to observe, with the first reliable discoveries made only in the mid-1990s \citep{first_BD,1995Natur.377..129R,1996ApJ...458..600B}. While hundreds of brown dwarfs have since been identified, the difficulty of their detection has largely limited the known population to the closest and youngest brown dwarfs in the Milky Way. This limitation poses two major problems. First, current research has been focused on sources with near-solar metallicities and chemical compositions which are not representative of the early evolutionary history of the Milky Way. Second, most of the ``evolved'' brown dwarfs currently known are isolated objects in the field which lack secondary indicators of their origins and physical properties, such as cluster membership or binary association. The theoretical challenges associated with modelling complex atmospheric chemistry and other low-temperature phenomena inhibits our ability to measure these physical properties accurately.

The population of brown dwarfs in globular clusters of the Milky Way addresses both of these problems. A typical globular cluster contains tens of thousands of individually observable coeval members with similar ages and chemical compositions that can be photometrically inferred from the colour-magnitude diagram of the population \citep{globulars}. The large masses of globular clusters allow their members to withstand tidal disruptions over extended periods of time, making these clusters some of the oldest coherent populations in the Milky Way ($\gtrsim10\ \mathrm{Gyr}$; \citealt{globular_ages,globular_ages_max}). In general, the long lifespans of globular clusters allow for extensive dynamical evolution: these gravitationally bound stellar systems tend towards thermodynamic equilibrium and energy equipartition, resulting in preferential segregation of members by mass and ejection of the lowest-mass stars and brown dwarfs \citep{globular_cluster_dynamics,gc_evaporation_1,gc_evaporation_2,gc_evaporation_3}. However, this effect is noticeably suppressed in the outer regions of globular clusters \citep{equilibrium_condition,no_equipartition}, whose relaxation times often exceed the age of the cluster \citep{gc_database} due to increased distances between the stars \citep[Ch. 2]{relaxation_time_equation}. These regions therefore preserve their primordial mass function and the mixing ratio between sub-populations within the cluster \citep{outskirt_IMF,outskirt_mixing,outskirt_evolution}.  

Unlike field stars in the solar neighbourhood, globular cluster members display chemical abundances characteristic of the early, metal-poor phases of the Milky Way's formation. 
Globular clusters are thus unique laboratories for studying brown dwarfs with non-solar abundances and old ages -- parameters that can be independently constrained from the overall cluster population. 
In turn, the abundance and cooling behavior of brown dwarfs make them potential instruments for refining the ages of host globular clusters \citep{age_gap,whitepaper,2004ApJS..155..191B}, in analogy to the use of brown dwarfs in age-dating young open clusters \citep{1998ApJ...499L.199S,2018ApJ...856...40M}.
Brown dwarfs thus provide a link between (sub)stellar evolution, galaxy formation and evolution, and cosmology (e.g., \citealt{age_universe_globular}).

The large distances to globular clusters and the faint luminosities of brown dwarfs have so far prevented the unambiguous detection of this distinct population. Existing deep photometric observations of Milky Way globular clusters, made primarily with instruments on the \textit{Hubble Space Telescope} (\textit{HST}), have reached the faint end of the Main Sequence \citep{M4_BD_limit,faint_end} and motivated dedicated searches for brown dwarfs in the nearest systems \citep{BD_hunt_1,BD_hunt_2}, although results from the latter remain ambiguous. The upcoming generation of large ground and space-based observatories, such as the \textit{James Webb Space Telescope} (\textit{JWST}), the Thirty Meter Telescope (TMT), the Giant Magellan Telescope (GMT), and the Extremely Large Telescope (ELT),
are expected to change this situation in the next few years \citep{JWST_proposal_1,JWST_proposal_2}. The promise of observational data for globular cluster brown dwarfs necessitates development of a theoretical framework for characterizing these sources, in particular evolutionary tracks and model atmospheres across the brown dwarf limit for non-solar abundances. 

In this work, we evaluate current \textit{HST} data and make predictions for forthcoming \textit{JWST} data for one of the most well-studied globular clusters in the Milky Way: \omegacen{} \citep{omegacen_discovery_1,omegacen_discovery_2}.
This system is the largest known globular cluster ($4\times10^6\ \mathrm{M}_\odot$, $10^7$ members; \citealt{omega_cen_mean_mass,omega_cen_total_mass}) and its dynamics fall far short of complete energy equipartition, as confirmed by direct measurements of the velocity distribution \citep{omega_cen_equipartition_coefficient} and constraints on mass segregation \citep{omega_cen_equipartition_segregation}. Our analysis is based on a sample located at $3$ half-light radii away from the cluster center where the relaxation time reaches $\sim4\times10^{10}\ \mathrm{Gyr}$ \citep{omega_cen_relaxation}, indicating a nearly pristine primordial population of brown dwarfs and low-mass stars.

\omegacen{} possesses two distinct populations, identified in a bifurcation of its optical Main Sequence into ``blue'' and ``red'' sequences \citep{anderson_thesis,bedin_bifurcation}. 
Away from the center of the cluster, the red sequence of \omegacen{} is the dominant population with over twice as many members as compared to the blue sequence \citep{discrete_populations_experimental}. Since metal-rich stars generally appear redder than their metal-poor counterparts due to heavier metal line blanketing at short wavelengths \citep{metal_reddening}, a top-heavy metallicity distribution in \omegacen{} is naively expected. However, this expectation is at odds with earlier spectroscopy of individual bright stars (e.g. \citealt{omegacen_metals}) that indicated a bottom-heavy distribution in metallicity among cluster members. By comparing the observed bifurcation to model isochrones, \citet{bedin_bifurcation} determined that the colour-magnitude diagram is unlikely to be explained by the spread in metallicity alone, nor by a background object with different chemistry. It was further suggested that the blue sequence may have a higher metallicity than the red sequence if it is significantly helium-enhanced, with a helium mass fraction ($\mathrm{Y}$) in excess of $0.3$ \citep{bedin_bifurcation}. Higher helium content increases the mean molecular weight in stellar interiors, producing hotter and bluer stars for identical masses and ages.

Subsequent quantitative analysis in \citet{omega_cen_helium_difference} found the helium mass fraction discrepancy between the sequences to be $\Delta \mathrm{Y}\sim0.12$. A follow-up spectroscopic study of identified members of red and blue sequences in \citet{spectroscopic_helium} confirmed that the metallicity of blue sequence members indeed exceeds that of red sequence members by $\sim0.3\ \mathrm{dex}$, strongly favouring the helium enhancement hypothesis. Consistent with all aforementioned results, \citet{helium_039} calculated the helium mass fraction of the blue sequence as $\mathrm{Y}=0.39\pm0.02$ which remains the most accurate estimate to date (an analysis in \citealt{helium_alternative} based on a different selection of sequence members and a different set of evolutionary models suggests that this value may be overestimated by $\gtrsim 0.05$). The origin of such extraordinarily high helium content remains 
under debate \citep{discrete_populations,omega_cen_helium_difference,blackhole_helium}.
An additional noteworthy aspect of \omegacen{} members is the scatter in stellar metallicities within each of the two sequences, which is fairly wide compared to other globular clusters \citep{populations_2,omega_cen_MDF}. This scatter suggests that \omegacen{} may be the nucleus of a nearby dwarf galaxy accreted by the Milky Way; or it may be a system intermediate in scale between a dwarf galaxy and a globular cluster \citep{omega_cen_dwarf_galaxy,omega_cen_dwarf_galaxy_2,2014MNRAS.443.1151N}. Indeed, recent work employing ultraviolet and infrared photometry and benefiting from the enlarged colour baselines 
was able to show 
that the red and blue sequences are each composed of multiple stellar subgroups, totalling up to 15 distinct sub-populations 
\citep{populations_2}.

In this study, we calculate new interior and atmosphere models for ages and non-solar chemical compositions appropriate for the members of \omegacen{}. By comparing synthetic colour-magnitude diagrams (CMDs) inferred from those models to new \textit{HST} photometric observations of the low-mass Main Sequence ($\lesssim 0.5\ \mathrm{M}_\odot$), we determine best-fit physical properties of the cluster and calibrate for interstellar reddening. Finally, we extend our models into the substellar regime to make predictions of expected colours, magnitudes, and colour-magnitude space densities of brown dwarfs in \omegacen{} down to effective temperatures of $T_\mathrm{eff}\approx 1000\ \mathrm{K}$.
Section~\ref{sec:2_method} provides an overview of our approach to modelling the \omegacen{} stellar and substellar population.
Section~\ref{sec:3_isochrones} describes how synthetic isochrones for the members of \omegacen{} were calculated, including our choices of specific cluster properties such as age and metallicity. We also briefly examine the role of atmosphere-interior coupling in our evolutionary models and discuss the relation of atmospheric and core lithium abundance predicted by our framework. Section~\ref{sec:3a_observations} describes our astro-photometric observations of \omegacen{} with \textit{HST}. 
Section~\ref{sec:4_evaluation} presents our method of comparing the isochrones against our photometry, and corresponding constraints on the best-fit physical parameters. 
Section~\ref{sec:5_predictions} provides predictions of the observable properties of brown dwarfs in the cluster in the context of future \textit{JWST} observations. 
Section~\ref{sec:6_conclusion} summarizes our results. Appendix~\ref{sec:8_mesa} describes the parameters of evolutionary models calculated in this study.
Appendix~\ref{sec:7_appendix} lists our choices of standard solar abundances. Finally, Appendix~\ref{sec:7b_catalog} provides a description of the \textit{HST} dataset for \omegacen{} used in this study that is included as an associated dataset.

\section{Overview of Methodology} \label{sec:2_method}
For the modelling purposes of this study, we define a stellar population as a group of stars and brown dwarfs with identical age, initial chemical composition, and distance from the Sun. While allowing for multiple co-existing populations in \omegacen{}, we require all of them to be drawn from the same initial mass function (IMF). The reality of a continuous, rather than discrete, distribution of chemical abundances among the members of the cluster is partly accounted for by allowing statistical scatter in the colour-magnitude space (see Section~\ref{sec:4_evaluation}). Potential variations in age are briefly considered in Section~\ref{sec:5_predictions}.

Our first step was to determine the best-fitting isochrone to our optical and near infrared \textit{HST} observations of \omegacen{} (see Section~\ref{sec:3a_observations}) that capture most of the Main Sequence but are not sensitive enough to reach the substellar regime. 
The multiplicity of populations in \omegacen{} necessitated an approximate categorization of the cluster as a whole
due to the extreme computational demand associated with calculating complete grids of model atmospheres and interiors for multiple sets of chemical abundances.
We therefore made no attempt to model the observed blue and red sequences separately; instead, we sought to model the modal colour-magnitude trend of the entire cluster.
Due to the narrow colour separation between the two sequences along the stellar Main Sequence \citep{populations}, we expect the modal trend to predict the colours and magnitudes of brown dwarfs in \omegacen{} for both populations. 

We started with an initial estimate of chemical abundances based on photometric and spectroscopic analysis of bright members in the literature \citep{abundances}. The helium abundance was set to the value corresponding to the blue sequence of the cluster from \citet{helium_039}. As will be demonstrated shortly, the enhanced helium mass fraction in combination with freely varying metal abundances results in a population that provides a satisfactory approximation of the modal colour-magnitude trend for \textit{both} red and blue sequences. On the other hand, we found the mass-luminosity relation of the cluster to be far more sensitive to the helium mass fraction such that no modal population could be obtained that would adequately fit the mass-luminosity relations of both red and blue sequences (see Section~\ref{sec:4_evaluation}). We therefore chose to adopt a distinct mass-luminosity relation for the red sequence from the literature \citep{DSED} and focus our new calculations on the helium abundance of the blue sequence. This choice was made for two reasons: first, due to the scarcity of helium-enhanced stellar models in the literature; and second, because higher helium content generally results in higher luminosities for the largest-mass brown dwarfs (e.g. compare models B and G in \citealt{burrows_models}; see also \citealt{1993RvMP...65..301B,BD_mass_radius,burning_limit_helium_2}). The latter effect makes helium-enriched brown dwarfs more likely to be detected in future magnitude-limited surveys.

We refer to the population based on this initial set of abundances as the \textit{nominal population} of the cluster. A synthetic isochrone was calculated and compared to existing photometry, and the chemical abundances of the nominal population (with the exception of helium) were perturbed iteratively until a best quantitative fit to the modal colour-magnitude trend of the cluster was obtained. We refer to all perturbed populations as \textit{secondary populations}. In line with our simplified model, we assumed that the entire CMD of the cluster could be described with one modal population, with an empirically determined scatter used to account for other sub-populations, multiple star systems, and measurement uncertainty.

Next, we sought to reproduce the observed present-day luminosity function (LF) of the cluster by combining the mass-luminosity relation of the best-fitting isochrone with the commonly used broken power law IMF (e.g. \citealt{Kroupa,omega_cen_IMF,broken_power_ex_1,broken_power_law_ex_2}). As explained above, we adopted an additional solar helium mass-luminosity relation from literature \citep{DSED} and added the contributions of both populations together in the LF using a population mixing ratio optimized through fitting. As demonstrated in Section~\ref{sec:4_evaluation}, a reasonably good match to the observed LF can be obtained with a simple two-component IMF and two stellar populations. 
Finally, the isochrone of the calculated best-fit population and the determined IMF were extended into the substellar regime to make predictions for the colours and magnitudes of brown dwarfs expected to be identified by {\em JWST}.

The isochrones and mass-luminosity relations for the nominal and secondary populations were calculated from corresponding grids of newly computed model atmospheres and interiors. Simultaneous coupled modelling of atmospheres and interiors is challenging, as the substantial difference in physical conditions between the two requires distinct numerical approaches typically implemented in independent software packages. In addition, atmosphere modelling tends to be orders of magnitude more computationally demanding, largely due to the complex molecular chemistry and opacity present at low temperatures. For those reasons, we followed the standard approach \citep{BCAH97,MIST} in which a grid of model atmospheres is pre-computed, covering the regions of the parameter space the stars are expected to encounter during their evolution.
To assure that the size of the model grid was computationally feasible, we restricted the number of degrees of freedom that are allowed to vary from atmosphere to atmosphere within the same population. The atmosphere grid for each population has been calculated over a range of effective temperatures ($T_\mathrm{eff}$) and surface gravities ($\log_{10}(g)$) encompassing the evolutionary states of low-mass stars and brown dwarfs, while all other parameters were assumed fixed across the population (e.g. elemental abundances, age) or derivable from the grid parameters (e.g. stellar radius). A synthetic spectrum was calculated for each model atmosphere in the grid, which could be subsequently converted to synthetic photometry for instruments of interest.

\section{Isochrones} \label{sec:3_isochrones}
\subsection{Initial parameters}

The parameters adopted for the nominal population are listed in Table~\ref{table:fixed_properties}. All abundances are given with respect to their standard solar values summarized in Appendix~\ref{sec:7_appendix}.

\begin{deluxetable}{lr|rr}
\tablenum{1}
\tablecaption{Properties of nominal population \label{table:fixed_properties}}
\tablewidth{0pt}
\tablehead{
\colhead{Parameter} & \colhead{} & \colhead{} & \colhead{Value}
}
\startdata
   Metallicity & $[\rm{M}/\rm{H}]$ & $-1.7$ & dex over solar  \\  \hline
   Carbon abundance & $[\rm{C}/\rm{M}]$ & $-0.65$ & dex over solar   \\ \hline
   Nitrogen abundance & $[\rm{N}/\rm{M}]$ & $1.45$ & dex over solar   \\ \hline
   Oxygen abundance & $[\rm{O}/\rm{M}]$ & $-0.1$ & dex over solar   \\ \hline
   Age & {} & 13.5 & $\rm{Gyr}$   \\ \hline
   Helium mass fraction& $\rm{Y}$ & $0.4$  &  \\ \hline
   Atmospheric lithium & $[\rm{Li}/\rm{M}]$ & $-3.0$ & dex over solar   \\ \hline
\enddata
\end{deluxetable}

\begin{figure}[h]
    \centering
    \includegraphics[width=1\columnwidth]{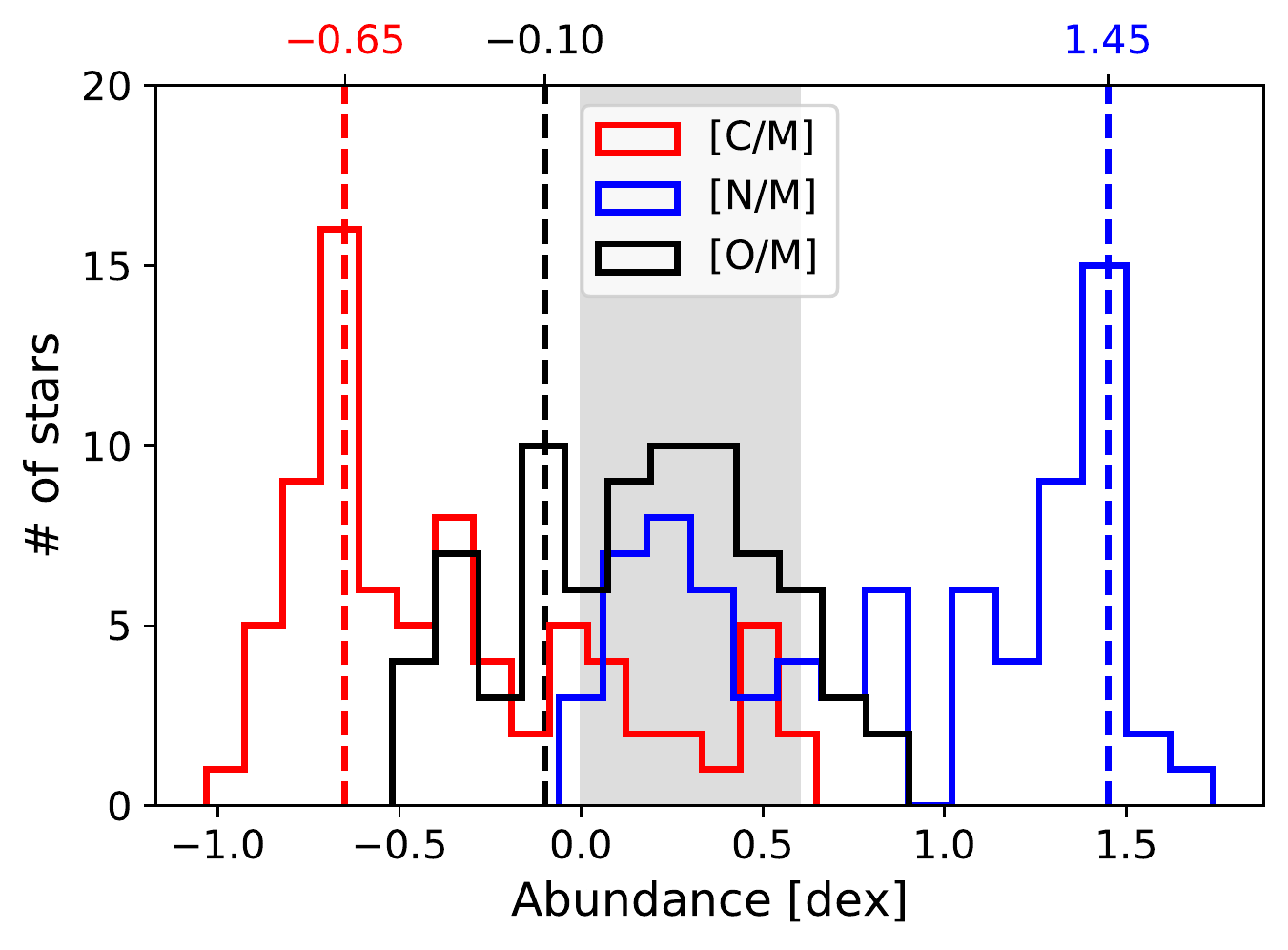}
    \caption{Distribution of measured elemental abundances from individual spectroscopy of 77 bright members of \omegacen{} from \citet{abundances}. The vertical dashed lines represent the values adopted in this study for the nominal population as per Table~\ref{table:fixed_properties}. The shaded area represents the range of oxygen abundances considered in secondary populations as per Table~\ref{table:table2}.}
    \label{fig:abundance_hists}
\end{figure}

The abundances of carbon ($[\rm{C}/\rm{M}]$), nitrogen ($[\rm{N}/\rm{M}]$) and oxygen ($[\rm{O}/\rm{M}]$) were selected to approximate the modes of the distributions inferred from individual spectroscopy of 77 bright ($10.3 < I < 12.7$) cluster members from \citet{abundances}. These distributions are shown in Figure~\ref{fig:abundance_hists}. Contrary to carbon and nitrogen, oxygen abundance lacks a well-defined modal peak and appears to vary in the range $-0.1 \lesssim [\rm{O}/\rm{M}] \lesssim 0.6$. For the nominal population, we chose the lower bound of the oxygen distribution in the figure since the data from \citet{abundances} suggest a correlation between $[\rm{C}/\rm{M}]$ and $[\rm{O}/\rm{M}]$, with the debiased Pearson coefficient of $0.72\pm0.03$; and an anti-correlation between $[\rm{N}/\rm{M}]$ and $[\rm{O}/\rm{M}]$ with the coefficient of $-0.61\pm0.04$. The lower bound on $[\rm{O}/\rm{M}]$ is therefore consistent with the modal peaks in $[\rm{C}/\rm{M}]$ and $[\rm{N}/\rm{M}]$ that appear to fall close to the low and high bounds of their corresponding distributions respectively. We note that the choices made for the nominal population are less important, as a secondary population will be used in the final analysis that best fits the data.

\begin{figure}[h]
    \centering
    \includegraphics[width=1\columnwidth]{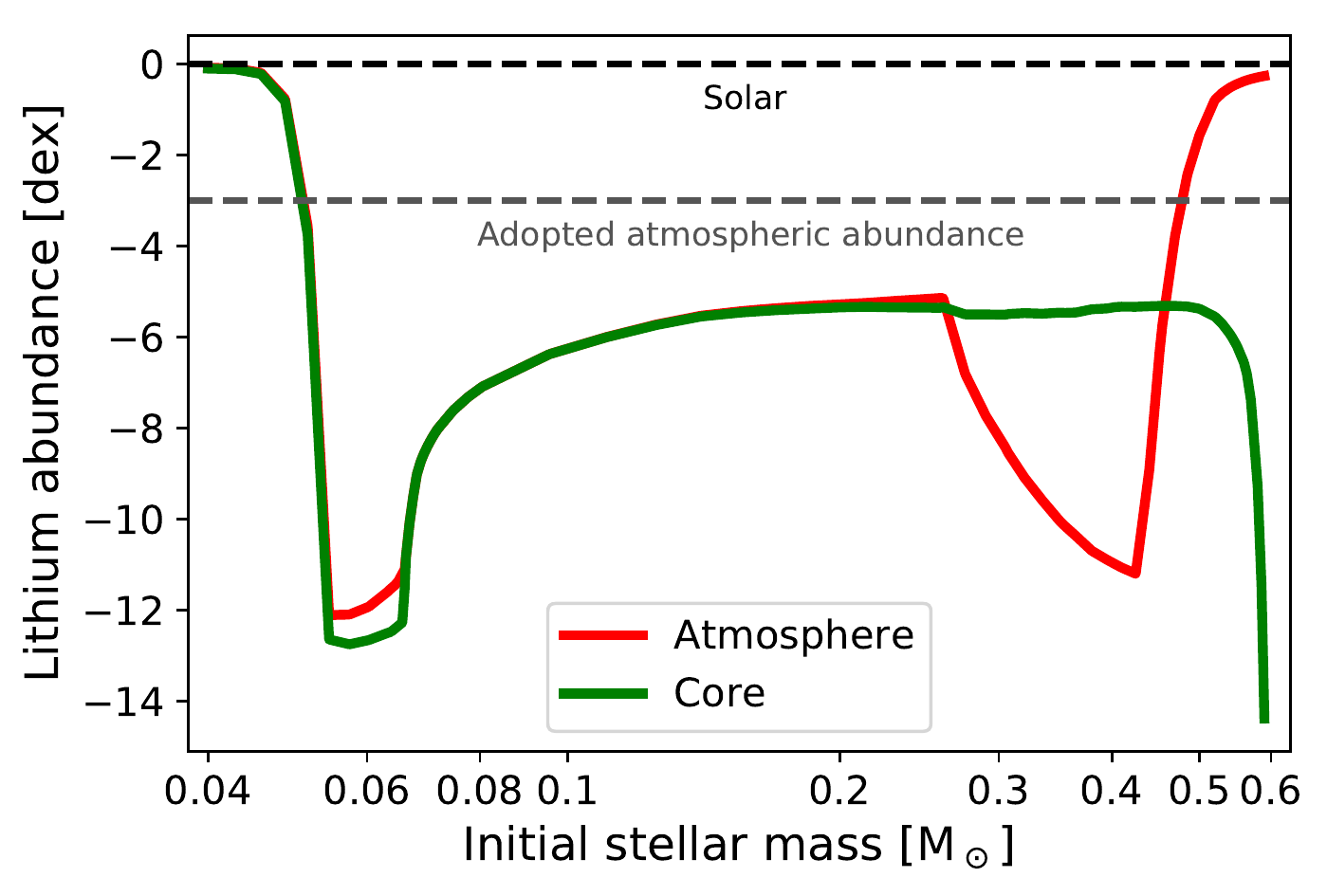}
    \caption{Depletion of lithium in the core and the atmosphere as a function of stellar mass for the \texttt{HMMA} secondary population (see Table~\ref{table:table2}) over $13.5\ \mathrm{Gyr}$. All models are initialized with a solar lithium abundance (see Appendix~\ref{sec:7_appendix}) in the PMS phase. Atmospheric lithium is not depleted at $M\lessapprox0.055\ \mathrm{M}_\odot$ due to insufficiently high temperatures for fusion, and at $M\gtrapprox0.5\ \mathrm{M}_\odot$ due to the early formation of a radiative zone that ``freezes'' the surface abundance. At intermediate masses, lithium is depleted by proton capture in the core which is propagated into the atmosphere via convective mixing. At masses above $\approx 0.07\ \mathrm{M}_\odot$, trace amounts of lithium are also produced by incomplete proton-proton chains. For masses below $\approx 0.3\ \mathrm{M}_\odot$ no radiative zone exists and lithium abundances are nearly equally depleted throughout the star. A radiative zone forms between $0.3\ \mathrm{M}_\odot$ and $0.4\ \mathrm{M}_\odot$, where the atmospheric abundance first decreases compared to core due to late formation of the radiative zone and then increases due to early formation. A late radiative zone allows lithium depletion by proton capture to propagate into the envelope but prevents diffusion of lithium enhancement from the proton-proton chain.}
    \label{fig:lithium}
\end{figure}

For every population, two sets of elemental abundances must be chosen: one for the zero age pre-main-sequence star (PMS) which will be used in evolutionary interior models; and one for the corresponding grid of model atmospheres. Ideally, the latter set must be informed by the final stages of fully evolved stars calculated using the former set. Unfortunately, this approach is not compatible with our method, in which the grid of model atmospheres is computed before the evolutionary models, necessitating an approximate treatment. With the exception of lithium, we assumed that the final atmospheric abundances match the initial PMS abundances, since any changes in composition induced by core nuclear fusion are expected to be insignificant at low masses, while models of higher mass ($\gtrsim 0.3\ \mathrm{M}_\odot$) develop interior radiative zones that preserve PMS abundances in the outer layers.
Our calculated evolutionary models (to be described below)
affirm this choice, with changes in abundances other than $\mathrm{Li}$ between the PMS and the surface of the fully evolved star never exceeding $\sim 0.1\ \mathrm{dex}$. On the other hand, the variation in lithium abundance in both the core and the atmosphere is significant, as shown in Figure~\ref{fig:lithium}. Atmospheric lithium is almost entirely consumed through proton capture for all but the smallest mass (insufficient central temperature for $\mathrm{Li}$ fusion) and the largest mass (formation of a radiative zone) models. Due to the minimal effect of lithium abundance on the stellar spectrum (and, even more so, synthetic photometry), we chose to ignore the minority of masses where $\mathrm{Li}$ is not depleted and assumed an abundance of $\mathrm{[Li/M]}=-3.0$ for all model atmospheres (but not for PMS in evolutionary models). This choice effectively eliminates lithium from the spectra.

The overall metallicity of the nominal population was chosen following \citet{populations}, who fit model isochrones onto \omegacen{} photometry acquired with the \textit{HST} 
\textit{Wide Field Channel} of the \textit{Advanced Camera for Surveys (ACS/WFC)} 
\citep{ACS_handbook} and 
the \textit{Infra Red} channel 
of the \textit{Wide Field Camera 3 (WFC3/IR)}
\citep{WFC3_handbook}. While the isochrones in \citet{populations} do not account for non-solar CNO abundances, they were consistent with observations and thus provide satisfactory starting parameters. Of the stellar populations identified in \citet{populations}, we specifically focused on the metal-poor side ($[\rm{Fe}/\rm{H}]\approx[\rm{M}/\rm{H}]\gtrapprox -1.7$) of the helium-rich ($\rm{Y}\approx 0.4$) MS-II population that corresponds to the blue sequence in \citet{bedin_bifurcation}. We set the lowest metallicity in the quoted range of MS-II as the initial guess for the nominal population, and allowed it to increase up to $[\rm{M}/\rm{H}]=-1.4$ in the secondary populations. We fixed the helium mass fraction to $\textrm{Y}=0.4$ for both nominal and secondary populations in accordance with both \citet{helium_039} and \citet{populations}.

\citet{populations} chose an isochrone age of $13.5\ \rm{Gyr}$, which we used in this investigation as well. The exact age of the cluster has little effect on the Main Sequence, which justifies using a single upper limit for the isochrone fitting regardless of the known variation in ages of individual members by a few $\rm{Gyr}$ \citep{abundances}. In contrast, brown dwarfs continuously evolve across colour-magnitude space, so our predictions were calculated for both $10\ \rm{Gyr}$ and $13.5\ \rm{Gyr}$ (Section~\ref{sec:5_predictions}).

Due to the multitude of populations in \omegacen{} and the inevitable bias in abundances inferred by individual stellar spectroscopy,
we perturbed the aforementioned parameters to generate $5$ sets of models for secondary populations, whose abundances are listed in Table \ref{table:table2}. The perturbations were applied iteratively until the best fit to the observed population was achieved (see Section~\ref{sec:4_evaluation}). All properties that are not mentioned in the table are identical to the nominal population. 

\begin{deluxetable}{lccc}
\tablenum{2}
\tablecaption{Properties of secondary populations \label{table:table2}}
\tablewidth{0pt}
\tablehead{
\colhead{Population} & \colhead{$[\mathrm{O/M}]$} & \colhead{$[\alpha\mathrm{/M}]$\tablenotemark{a}} & \colhead{$[\mathrm{M/H}]$}
}
\startdata
\texttt{LMHO} (\texttt{L}ow \texttt{M}etal \texttt{H}igh \texttt{O}xygen)   &    $0.6$ &    $0.0$  &    $-1.7$  \\ \hline
\texttt{HMET} (\texttt{H}igh \texttt{MET}al)   &    $0.0$ &    $0.0$  &    $-1.4$  \\ \hline
\texttt{HMMO} (\texttt{H}igh \texttt{M}etal \texttt{M}edium \texttt{O}xygen)   &    $0.4$ &    $0.0$  &    $-1.4$  \\ \hline
\texttt{HMMA} (\texttt{H}igh \texttt{M}etal \texttt{M}edium \texttt{A}lpha)   &    $0.0$ &    $0.4$  &    $-1.4$  \\ \hline
\texttt{HMHA} (\texttt{H}igh \texttt{M}etal \texttt{H}igh \texttt{A}lpha)   &    $0.0$ &    $0.6$  &    $-1.4$  \\ \hline
\enddata
\tablenotetext{a}{$[\alpha\mathrm{/M}]$ refers to the enhancement of $\alpha$-elements that include $\mathrm{O}$, $\mathrm{Ne}$, $\mathrm{Mg}$, $\mathrm{Si}$, $\mathrm{S}$, $\mathrm{Ar}$, $\mathrm{Ca}$ and $\mathrm{Ti}$}
\end{deluxetable}

\begin{deluxetable}{cp{0.6\columnwidth}r}
\tablenum{3}
\tablecaption{Molecular lines included in our \texttt{PHOENIX} setup \label{table:molecular}}
\tablewidth{0pt}
\tablehead{
\colhead{Ref} & \colhead{Molecules} & \colhead{\# of lines}
}
\startdata
(1) & $\mathrm{H}\mathrm{O}\mathrm{D}$   &   $41.3\times 10^6$ \\ \hline
(2) & $\mathrm{H}_2\mathrm{O}$   &   $505\times 10^6$ \\ \hline
(3) & $\mathrm{CC}$, $\mathrm{CN}$, $\mathrm{CH}$, $\mathrm{NH}$, $\mathrm{OH}$, $\mathrm{SiO}$, $\mathrm{SiH}$, $\mathrm{H}_2$ &   $5.7\times 10^6$ \\ \hline
(4) & $\mathrm{CO}_2$ &   $4\times 10^6$ \\ \hline
(5) & $\mathrm{NH}_3$ &   $6.7\times 10^3$ \\ \hline
(6) & $\mathrm{ZrO}$, $\mathrm{YO}$ &   $267\times 10^3$ \\ \hline
(7) & $\mathrm{CO}$ &   $134\times 10^3$ \\ \hline
(8) & $\mathrm{C}_{2}\mathrm{H}_{2}$, $\mathrm{C}_{2}\mathrm{H}_{4}$, $\mathrm{C}_{2}\mathrm{H}_{6}$, $\mathrm{COF}_{2}$, $\mathrm{CH}_{3}\mathrm{OH}$, $\mathrm{CH}_{3}\mathrm{D}$, $\mathrm{N}_{2}$, $\mathrm{N}_{2}\mathrm{O}$, $\mathrm{NO}$, $\mathrm{NO}_{2}$, $\mathrm{NH}_{3}$, $\mathrm{OCS}$, $\mathrm{O}_{2}$, $\mathrm{O}_{3}$, $\mathrm{SO}_{2}$, $\mathrm{SF}_{6}$, $\mathrm{HI}$, $\mathrm{HCN}$, $\mathrm{HCOOH}$, $\mathrm{HNO}_{3}$, $\mathrm{HOCl}$, $\mathrm{HOBr}$, $\mathrm{HO}_{2}$, $\mathrm{HOD}$, $\mathrm{HF}$, $\mathrm{HCl}$, $\mathrm{HBr}$, $\mathrm{H}_{2}\mathrm{CO}$, $\mathrm{H}_{2}\mathrm{O}_{2}$, $\mathrm{H}_{2}\mathrm{O}$, $\mathrm{H}_{2}\mathrm{S}$  &   $1.3\times 10^6$ \\ \hline
(9) & $\mathrm{CO}_2$, $\mathrm{OH}$, $\mathrm{PH}_3$ &   $31.2\times 10^3$ \\ \hline
(10) & $\mathrm{CN}$ &   $2.2\times 10^6$ \\ \hline
(11) & $\mathrm{CH}_4$ &   $34.6\times 10^3$ \\ \hline
(12) & $\mathrm{H}_3^+$ &   $3.1\times 10^6$ \\ \hline
(13) & $\mathrm{CrH}$, $\mathrm{FeH}$, $\mathrm{TiH}$ &   $301\times 10^3$ \\ \hline
(14) & $\mathrm{MgH}$ &   $53.8\times 10^3$ \\ \hline
(15) & $\mathrm{CaH}$, $\mathrm{TiH}$, $\mathrm{VO}$ &   $14.6\times 10^6$ \\ \hline
(16) & $\mathrm{CH}_4$ &   $31.3\times 10^6$ \\ \hline
\enddata
\tablecomments{(1) -- AMES water \citep{water}, (2) BT water \citep{PL_BT}, (3) -- Kurucz CD-ROM \#15 \citep{PL_KCD15}, (4) -- \textit{CDSD} (Carbon Dioxide Spectroscopic Databank) \citep{PL_CDSD}, (5) -- \citet{PL_CS_HITRAN}, (6) -- \citet{PL_DRA}, (7) -- \citet{PL_GOORVITCH}, (8) -- \textit{HITRAN2004} \citep{PL_HITRAN2004}, (9) -- \textit{HITRAN2008} \citep{PL_HITRAN2008}, (10) -- \citet{PL_JOERGENSEN}, (11) -- \citet{PL_LB_HITRAN}, (12) -- \citet{PL_NT}, (13) -- \textit{MoLLIST} \citep{PL_BERNATH}, (14) -- \citet{PL_PHILLIP2}, (15) -- lines inherited from \texttt{MARCS} atmospheres \citep{PL_PLEZ2004}, (16) -- methane lines generated using \textit{STDS} (Spherical Top Data System; \citealt{PL_STDS1})  in \citet{PL_STDS2}.}
\end{deluxetable}

\subsection{Model atmospheres}

We calculate all model atmospheres with $T_\mathrm{eff}\leq 4000\ \mathrm{K}$ using a custom setup based on a branch of version 15.5 of the \texttt{PHOENIX} code \citep{phoenix_origin}. 
Molecular lines considered in the calculation are listed in Table~\ref{table:molecular}.
Our modelling framework includes the formation of condensate clouds in the atmosphere and their depletion by gravitational settling according to the \textit{Allard \& Homeier} cloud formation model \citep{phoenix_15,cloud_model_comparison}.  At $T_\mathrm{eff}<3000\ \mathrm{K}$ we used the ``cloudy'' mode described in \citet{roman_note}. For optimization purposes, a slightly simplified ``dusty'' mode is used at $T_\mathrm{eff}\geq 3000\ \mathrm{K}$, which differs in its coarser stratification ($128$ spherically symmetric layers instead of $250$), disabled gravitational settling, and fewer spectral features included in the calculation. It was verified that the transition between the two modes does not introduce noticeable discontinuities in the derived bolometric corrections and the difference between ``cloudy'' and ``dusty'' spectra at the transition temperature is insignificant. All \texttt{PHOENIX} models were calculated at wavelengths from $1\ \mbox{\normalfont\AA}$ to $1\ \mathrm{mm}$ with a median resolution of $\lambda/\Delta \lambda\approx18250$ in the range $\SI{0.4}{\micro\metre}\leq\lambda\leq \SI{2.6}{\micro\metre}$ and a lower resolution of $\sim 8000$ elsewhere.

At $T_\mathrm{eff}> 4000\ \mathrm{K}$, the effects of both condensates and molecular opacities become subdominant, allowing us to replace \texttt{PHOENIX} with the much faster and simpler \texttt{ATLAS} code version 9 \citep{ATLAS5,ATLAS_Linux, 9_vs_12, ATLAS_howto}.
As opposed to \texttt{PHOENIX}, our \texttt{ATLAS} setup stratifies the atmosphere into $72$ plane-parallel layers covering the range of optical depths from $\tau=100$ to $\tau \sim 10^{-7}$. Instead of direct opacity sampling, \texttt{ATLAS} relies on pre-computed opacity distribution functions (ODFs) \citep{Carbon}. Convection is modelled using mixing-length theory \citep{MLT,more_convection} with no overshoot. Modelled line opacities include $\sim 43\times10^6$ atomic transitions of various ionization stages and $\sim 123\times10^6$ molecular transitions including titanium oxide lines from \citet{schwenke_1998} and water lines from \citet{water}. We use satellite utilities \texttt{DFSYNTHE} and \texttt{SYNTHE} shipped with the main \texttt{ATLAS} code to compute a custom set of ODFs for the abundances of interest (one set for each considered population) and derive high-resolution synthetic spectra from the calculated models respectively. The calculated ODFs account for flux from $\sim10\ \mathrm{nm}$ to $\SI{160}{\micro\metre}$ to ensure correct evaluation of energy equilibrium through the atmosphere. On the other hand, our synthetic spectra span a narrower range of wavelengths from $\SI{0.1}{\micro\metre}$ to $\SI{4.2}{\micro\metre}$, accommodating all instrument bands considered in this study. All \texttt{SYNTHE} spectra are calculated at the resolution of $\lambda/\Delta\lambda=6\times10^5$.

\begin{figure*}
    \centering
    \includegraphics[width=2\columnwidth]{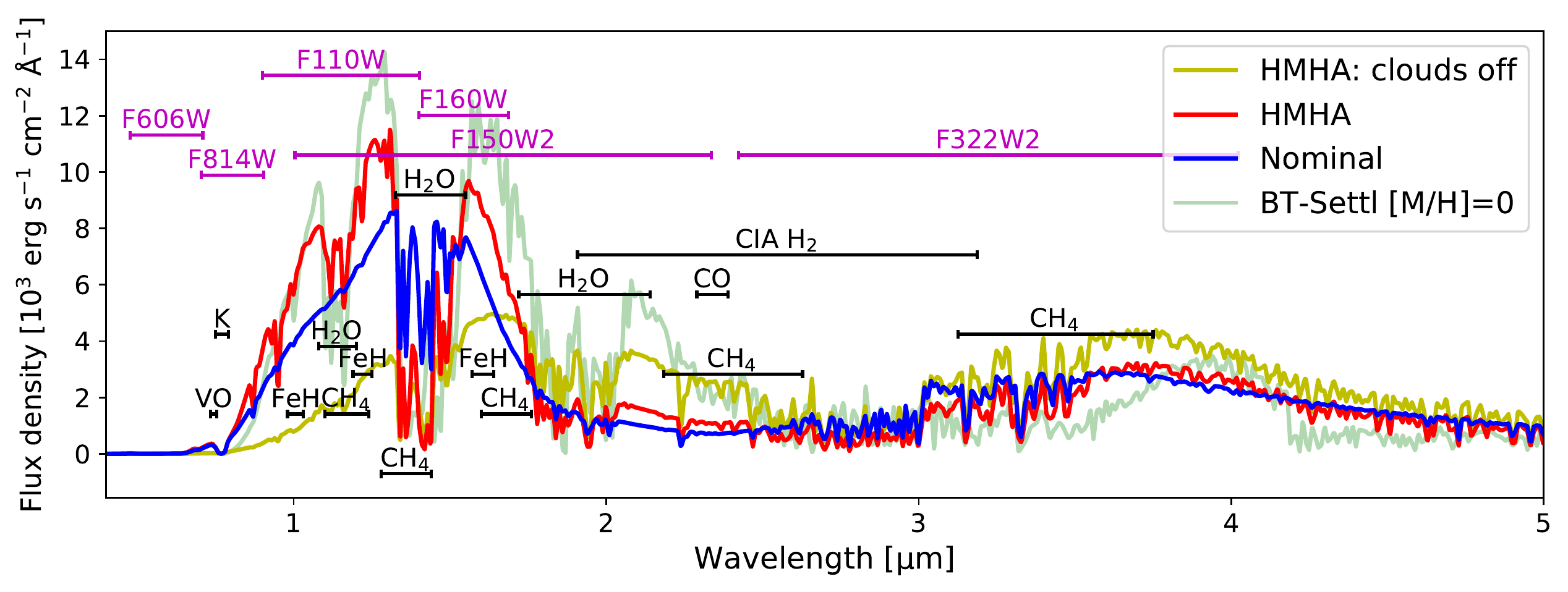}
    \caption{Synthetic spectra of selected low-temperature model atmospheres calculated in this study. Shown here are the $T_\mathrm{eff}=1200$, $\log_{10}(g)=5.0$ atmospheres from the nominal population (Table~\ref{table:fixed_properties}) and the \texttt{HMHA} population (Table~\ref{table:table2}). Both spectra demonstrate prominent molecular features, some of which are indicated with black bars (CIA $\mathrm{H}_2$ represents the band of collision-induced absorption by molecular hydrogen). A \texttt{HMHA} spectrum with identical parameters but calculated in the ``dusty'' mode (no gravitational settling) is shown for comparison. The corresponding synthetic spectrum for a model of solar metallicity from the \texttt{BT-Settl} library is also shown. Magenta bars delineate $20\%$ transmission bounds of \textit{HST} \textit{ACS/WFC} \texttt{F814W} and \texttt{F606W} bands; \textit{HST} \textit{WFC3/IR} \texttt{F110W} and \texttt{F160W} bands; and \textit{JWST} \textit{NIRCam} \texttt{F150W2} and \texttt{F322W2} bands. For clarity, the spectra are shown after convolution with a $3\ \mathrm{nm}$-wide Gaussian kernel.}
    \label{fig:example_spectra}
\end{figure*}

A few examples of calculated low-temperature models are plotted in Figure~\ref{fig:example_spectra}. Compared to their solar metallicity counterparts, the spectra of metal-poor brown dwarfs are characterized by weaker molecular absorption (e.g., $\SI{3.5}{\micro\metre}$ methane band), more prominent collision-induced $\mathrm{H}_2$ absorption originating from deeper layers of the atmosphere, and extreme pressure broadening of alkali metal lines (e.g., K~I resonant line at $\SI{0.77}{\micro\metre}$). Synthetic spectra computed under our setup have previously demonstrated good correspondence with observations of candidate metal-poor brown dwarfs in the field \citep{schneider}. All calculated model atmospheres are publicly available in our online repository\footnote{\href{http://atmos.ucsd.edu/?p=atlas}{\url{http://atmos.ucsd.edu/?p=atlas}}}.

A typical \texttt{PHOENIX} model in the ``cloudy'' mode requires $\sim 150$ CPU hours to converge on the Comet cluster at the San Diego Supercomputer Center made available to us through the \textit{XSEDE} programme \citep{6866038}. ``Dusty'' models were a factor of two or three faster to compute, while \texttt{ATLAS} models only took approximately $1$ CPU hour each.

\subsection{Atmosphere-interior coupling}

\begin{figure}[h]
    \centering
    \includegraphics[width=1\columnwidth]{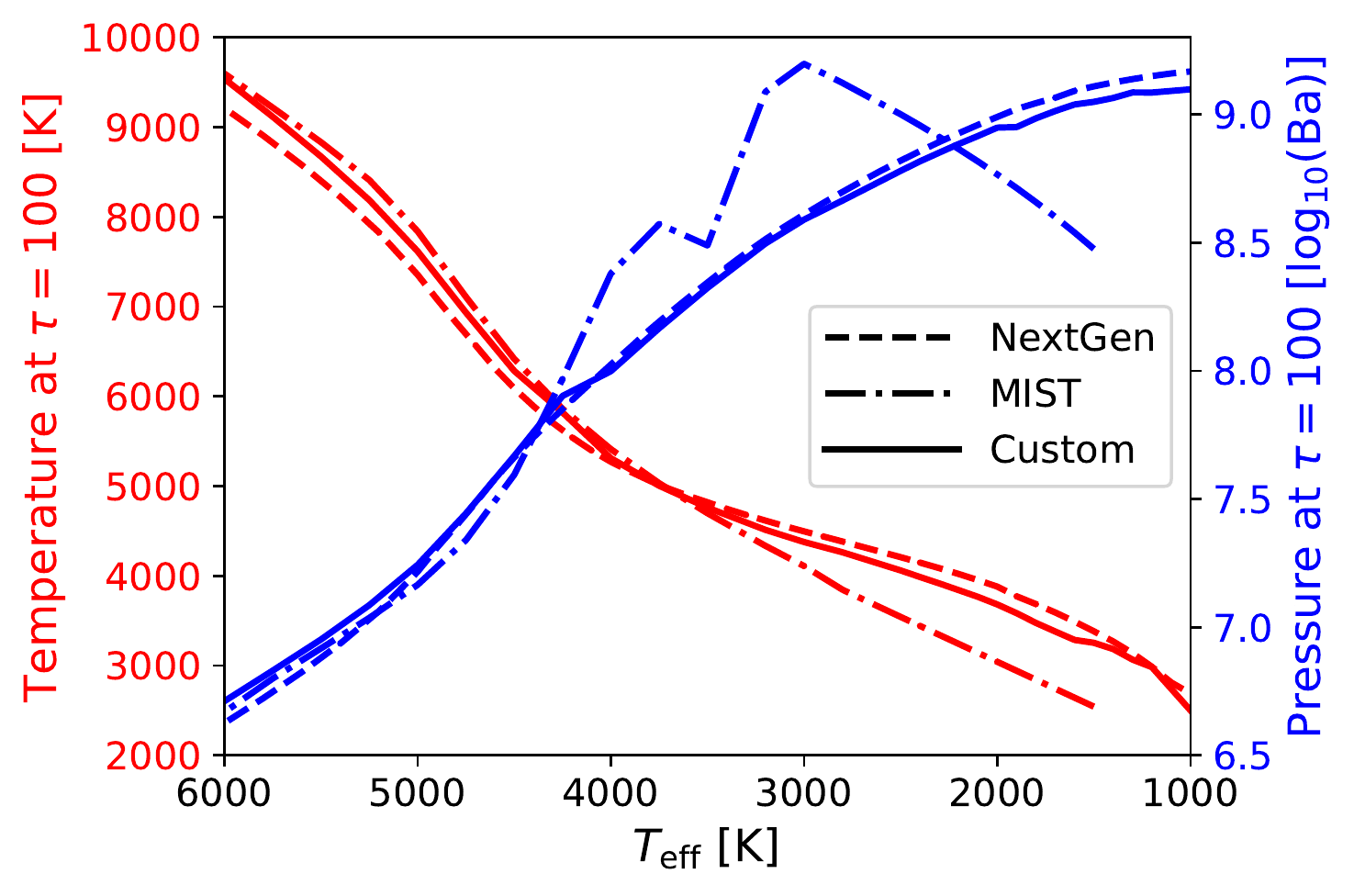}
    \caption{Comparison of three different sets of atmosphere-interior coupling boundary condition tables considered in this study at the surface gravity of $\log_{10}(g)=6.0$ and metallicity of $\mathrm{[M/H]}=-1.7$. \texttt{NextGen} (dashed line) refers to the \texttt{PHOENIX} grid from \citet{nextgen,nextgen_hot} that excludes gravitational settling in the atmosphere as well as enhancements of individual elements. \texttt{MIST} (dash-dotted line) refers to the \texttt{ATLAS}-derived tables used in \citet{MIST}. The ``custom'' coupling (solid line) is based on newly calculated \texttt{PHOENIX} models at low $T_\mathrm{eff}$ and \texttt{ATLAS} models at high $T_\mathrm{eff}$ and includes individual elemental enhancements of the nominal population (Table~\ref{table:fixed_properties}) in addition to the metallicity scaling as described in text. Pressure is shown in CGS units of barye ($1\ \mathrm{Ba}=1\ \mathrm{dyn}\ \mathrm{cm}^{-2}$).}
    \label{fig:boundary_tp}
\end{figure}

\begin{figure}[h]
    \centering
    \includegraphics[width=1\columnwidth]{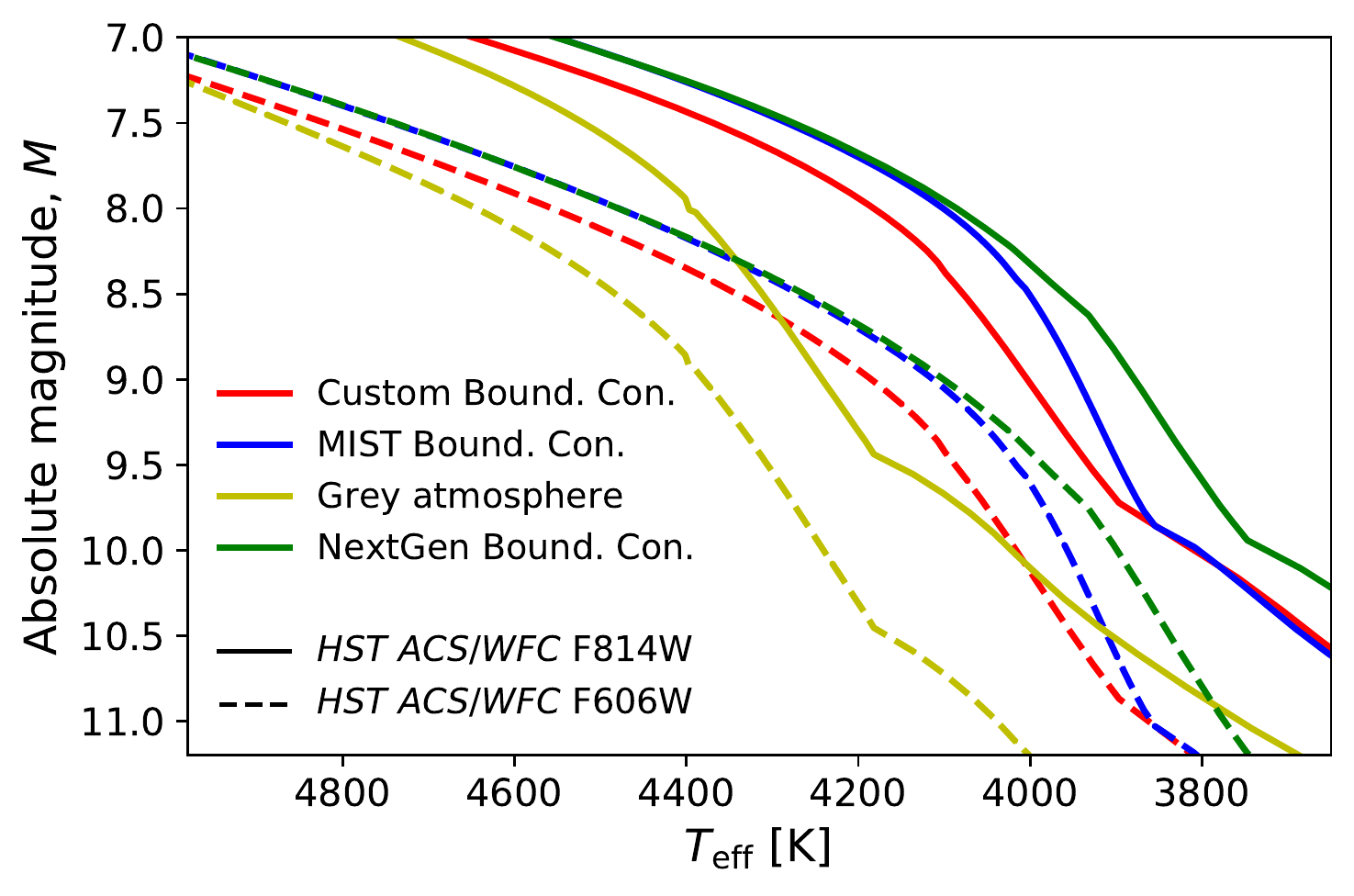}
    \caption{Effect of the choice of approach to atmosphere-interior coupling on synthetic photometry. The curves represent expected absolute magnitudes of the nominal population (see Table~\ref{table:fixed_properties}) as a function of effective temperature in two of the \textit{HST} \textit{ACS/WFC} bands: \texttt{F814W} (solid lines) and \texttt{F606W} (dashed lines), without interstellar extinction. The coupling schemes with pre-tabulated boundary conditions (Bound. Con.) are identical to those in Figure~\ref{fig:boundary_tp}. The grey atmosphere coupling scheme at $\tau=2/3$ is shown for comparison, which is the default scheme in \texttt{MESA}.}
    \label{fig:boundary_mag}
\end{figure}

We used the \texttt{MESA} code (Modules for Experiments in Stellar Astrophysics; \citealt{MESA}) for all evolutionary calculations. At zero age, a \texttt{MESA} model is spawned as a PMS with a given total mass and uniform elemental abundances. The initial structure is determined by assuming a fixed central temperature well below the nuclear burning limit (in our case, $5\times10^5\ \rm{K}$; \citealt{MIST}) and searching for a solution to the structure equations that reproduces the desired mass of the star. From here, evolution proceeds in dynamically determined time steps until the age of the model reaches the target age. On each step, the structure equations are solved using the atmospheric temperature and pressure as boundary conditions. Both can in principle be estimated from the current surface gravity and effective temperature of the model using an appropriate model atmosphere. It is those boundary conditions that establish the coupling between interiors and atmospheres. Once the interior structure of the star is known, the model can be advanced to the next time step by compounding expected changes due to diffusion, gravitational settling, nuclear reactions, mechanical expansion, and other time-dependent processes.

Our \texttt{MESA} configuration is derived from \citet{MIST} with a number of key differences outlined in detail in Appendix~\ref{sec:8_mesa}. When handling atmosphere-interior coupling, \texttt{MESA} is able to estimate boundary conditions either by drawing them from a pre-computed table at a given optical depth or at run time using one of a variety of methods relying on simplifying assumptions such as grey atmosphere. The latter option is unlikely to be accurate at low effective temperatures where molecular opacities and clouds dominate the spectrum. The low-mass \texttt{MESA} setup employed by \citet{MIST} relies on boundary condition tables calculated at $\tau=100$ for a wide range of effective temperatures, surface gravities and metallicities. However, the accuracy of the tables at $T_\mathrm{eff}\lessapprox 3500\ \mathrm{K}$ is questionable, as they were derived from \texttt{ATLAS} atmospheres that fail to account for significant low-temperature effects such as condensation and molecular features.

In this study, we compared four different approaches to atmosphere-interior coupling:

\begin{itemize}
    \item Run time calculation assuming grey atmosphere and drawing temperature and pressure at $\tau=2/3$;
    \item $\tau=100$ tables from \citet{MIST} at the \omegacen{} metallicity, but not accounting for individual elemental enhancements or low-temperature atmospheric effects absent in \texttt{ATLAS} atmospheres;
    \item Custom $\tau=100$ tables drawn from \texttt{NextGen}, a publicly available \texttt{PHOENIX} grid \citep{nextgen,nextgen_hot} without condensates or gravitational settling. The grid covers the \omegacen{} metallicity, but not the individual elemental enhancements; and
    \item Custom $\tau=100$ tables drawn from our own atmosphere grids described above based on \texttt{ATLAS} at high temperatures and \texttt{PHOENIX} at low temperatures, including condensation and gravitational settling. The grids include all individual elemental enhancements for each of the considered populations.
\end{itemize}

The grids of model atmospheres calculated in this study span surface gravities from $\log_{10}(g)=4$ to $\log_{10}(g)=6$. At early ages ($\lesssim2\ \mathrm{Myr}$), stars and brown dwarfs may briefly experience surface gravities under $\log_{10}(g)=4$, falling outside of the calculated atmosphere grid. In such instances, the boundary conditions from \citet{MIST} were used instead. By applying random perturbations to those low-gravity boundary conditions, we established that their accuracy has a negligible effect on the final results.

The temperature and pressure at $\tau=100$ for the tabular options are plotted as functions of effective temperature in Figure~\ref{fig:boundary_tp} at $\log_{10}(g)=6.0$. The effect of the chosen boundary conditions on synthetic photometry (described below) is shown in Figure~\ref{fig:boundary_mag}. Both figures demonstrate good agreement between approaches at high effective temperatures, and increasing deviation at lower temperatures where atmosphere-interior coupling becomes important. The final set of interior models in our analysis use custom $\tau=100$ tables based on our own model atmospheres, which we believe to offer the highest accuracy. The comparison of different sets of boundary conditions is presented here to emphasize the importance of atmosphere-interior coupling and to demonstrate how significant changes in metallicity and elemental enhancements could be ``mimicked'' by inaccurate boundary conditions.

\subsection{Synthetic photometry}

Synthetic photometry of each modelled population of \omegacen{} was computed by first evaluating the bolometric corrections of each bandpass of interest for each of the calculated model atmospheres. The bolometric correction is defined as

\begin{equation}
    \mathrm{BC}_x=M_b - M_x=M_b + 2.5 \log_{10}\left( \frac{F_x}{F_x'} \right)
    \label{eq1}
\end{equation}

\noindent where $x$ is a given bandpass; $\mathrm{BC}_x$ is the bolometric correction for $x$ between the  absolute bolometric magnitude $M_b$ and the absolute magnitude in band $x$, $M_x$; $F_x$ is the total flux of the model through bandpass $x$; and $F_x'$ is the total flux of the reference object through bandpass $x$. We used the \texttt{VEGAMAG} system for all comparisons to \textit{HST} data and the \texttt{ABMAG} system for \textit{JWST} predictions. For \texttt{VEGAMAG}, we used the apparent spectrum of Vega in \citet{vegamag} as our reference. For \texttt{ABMAG}, the reference spectrum is defined to be a constant flux density per unit frequency of $\approx3631\ \mathrm{Jy}$ at all frequencies \citep{abmag}. Both $F_x$ and $F_x'$ are measured in \textit{photons} per unit time per unit area \citep{photon_counting} since all instruments of interest are photon-counting. $F_x$ (but not $F_x'$) is taken at the distance of $10\ \mathrm{pc}$. By introducing the stellar radius $R$ we can express $F_x$ in terms of surface flux, $\Phi_x$:

\begin{equation}
\mathrm{BC}_x=M_b + 2.5 \log_{10}\left( \frac{\Phi_x}{F_x'} \right)+ 5 \log_{10}\left( \frac{R}{10\ \mathrm{pc}} \right)
\label{eq2}
\end{equation}

\noindent Both $R$ and $M_b$ are dependent on the total luminosity of the model, $L$, which cannot be inferred from the model atmosphere on its own. For our purposes, $\mathrm{BC}_x$ must be re-expressed in terms of exclusively atmospheric parameters. The IAU definition of absolute bolometric magnitude \citep{IAU_mag} is

\begin{equation}
M_b=-2.5 \log_{10}(L/[1\ \mathrm{W}]) + \Delta
\label{eq3}
\end{equation}

\noindent with $\Delta=71.197425$. Substituting in equation (\ref{eq2}):

\begin{equation}
\mathrm{BC}_x = 2.5 \log_{10}\left(\frac{\Phi_x}{F_x'} \right) - 10 \log_{10}\left(\frac{T_\mathrm{eff}}{1000~K} \right) + C
\label{eq4}
\end{equation}

\noindent where $C = -30.88138$ is a constant evaluated as

\begin{equation}
C=\Delta-2.5 \log_{10}\left[\frac{4\pi\sigma (10\ \mathrm{pc})^2(1000\ \mathrm{K})^4}{1\ \mathrm{W}} \right]
\label{eqC}
\end{equation}

\noindent with $\sigma$ representing the Stefan-Boltzmann constant.

Finally, we rewrite the flux ratio, $\Phi_x/F_x'$, in terms of the synthetic \textit{energy} spectrum $\phi_\lambda$, reference \textit{energy} spectrum $f'_\lambda$ and the dimensionless transmission profile of $x$: $x_{\lambda}$: 

\begin{equation}
\frac{\Phi_x}{F_x'} =  \frac{\int_0^\infty \lambda \phi_\lambda x_\lambda 10^{-\frac{A_\lambda}{2.5}} d\lambda}{\int_0^\infty \lambda f_\lambda' x_\lambda d\lambda}
\label{eq5}
\end{equation}

\noindent In the case of \texttt{ABMAG} magnitudes, $f'_\lambda$ must be converted from constant flux density per unit frequency as

\begin{equation}
{f'_\lambda}^{(\mathrm{ABMAG})}=(3631\ \mathrm{Jy}) \frac{c}{\lambda^2}
\label{eqAB}
\end{equation}

\noindent where $c$ is the speed of light.
Note that both integrands in equation (\ref{eq5}) are multiplied by $\lambda$ to express the spectra in photon counts rather than units of energy. We have also introduced $A_\lambda$ -- the extinction law in units of magnitude as a function of wavelength $\lambda$. We used the extinction law from \citet{extinction} parameterized by the optical interstellar reddening, $E(B-V)$, and the total-to-selective extinction ratio, $R_V=A_V/E(B-V)$. We assumed $R_V=3.1$ throughout and allowed $E(B-V)$ to be a free parameter, as described in Section~\ref{sec:4_evaluation}.

For each of the modelled populations, a synthetic colour-magnitude diagram was constructed by calculating a grid of interior models with initial masses spanning from the lowest mass covered by the calculated model atmospheres ($0.03\ \mathrm{M}_\odot$ for the best-fit isochrone) to the highest mass compatible with our atmosphere-interior coupling scheme ($\sim0.5\ \mathrm{M}_\odot$). At higher masses, $\tau=100$ lies too deep in the atmosphere, requiring a change in the reference optical depth \citep{MIST} and potentially causing a numerical discontinuity in the calculated results. Since the upper mass limit of $0.5\ \mathrm{M}_\odot$ is sufficient to accommodate the vast majority of the available \textit{HST} photometry (see Section~\ref{sec:4_evaluation}), we chose to restrict our analysis to this upper mass limit, thereby avoiding the complexities of using multiple atmosphere-interior coupling schemes.

The bolometric corrections in the bands of interest were calculated as described above for each model atmosphere in the grid. Due to convergence issues associated with cloud formation at very low effective temperatures, a few models with maximum flux errors in radiative zones exceeding $10\%$ were excluded from the atmosphere model grid. The remaining grid was then interpolated in effective temperature and surface gravity to the final surface parameters of each evolutionary interior model at the target age. Finally, the interpolated bolometric corrections were combined with the bolometric magnitudes of each interior model to obtain the desired synthetic photometry.

\subsection{Results}

\begin{figure}[h!]
    \centering
    \includegraphics[width=1\columnwidth]{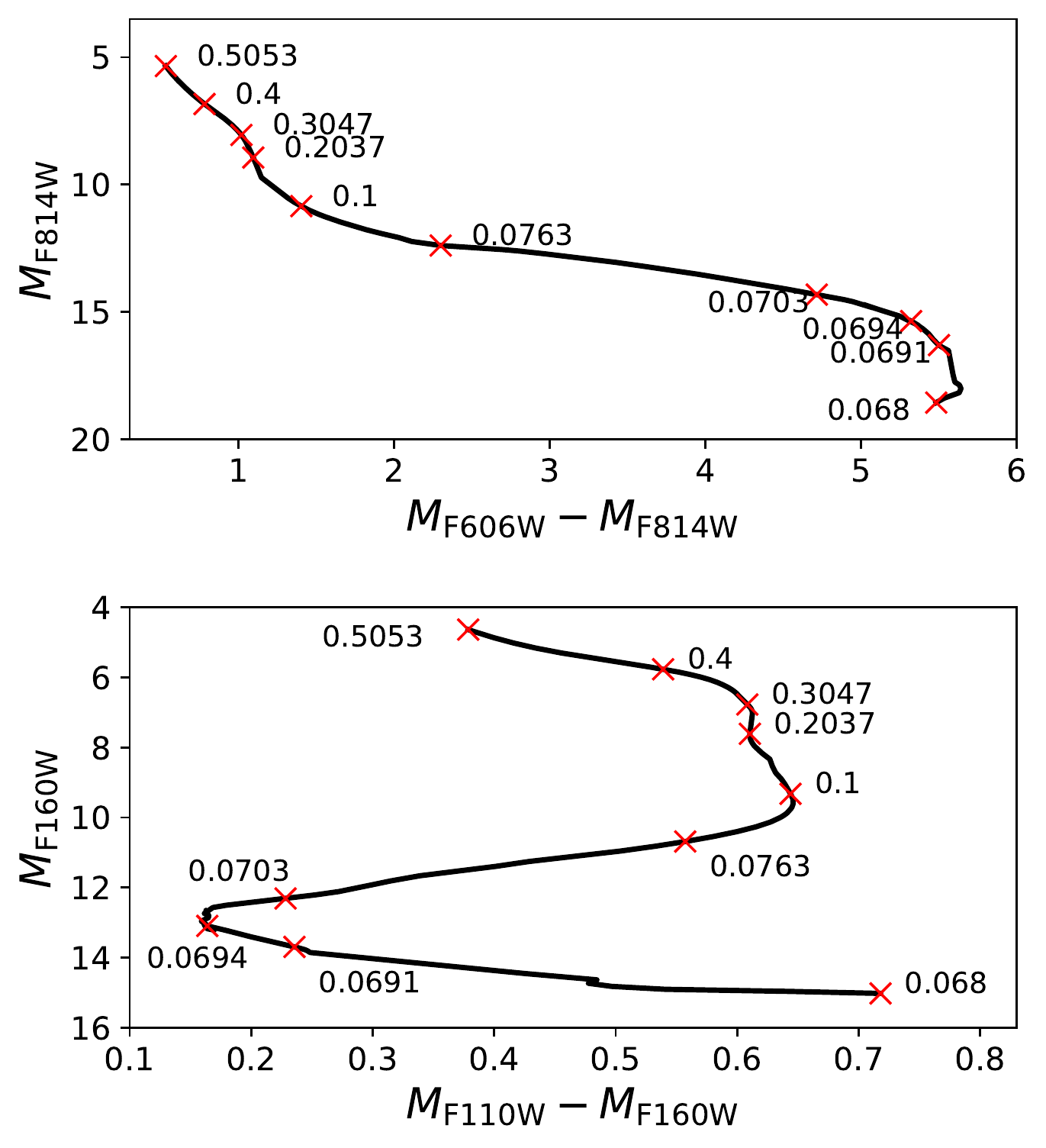}
    \caption{Isochrones derived for the nominal population of \omegacen{} in optical (\textit{top}) and near infrared (\textit{bottom}) absolute colour-magnitude spaces. The optical isochrone is evaluated for \textit{HST} \textit{ACS/WFC} filters, while the near infrared isochrone is evaluated for \textit{HST} \textit{WFC3/IR} filters. Red markers display the initial stellar masses of selected models along the isochrone in units of solar masses. Extinction effects are not included.}
    \label{fig:iso_nominal}
\end{figure}

Figure~\ref{fig:iso_nominal} shows the calculated isochrone of the nominal population of \omegacen{} as defined in Table~\ref{table:fixed_properties}. The isochrone is plotted in the \textit{absolute pre-extinction} colour-magnitude spaces defined by the \textit{HST} \textit{ACS/WFC} \texttt{F606W} and \texttt{F814W} optical bands and the \textit{HST} \textit{WFC3/IR} \texttt{F110W} and \texttt{F160W} near infrared bands. 
The isochrone displays a characteristic inflection point around $\sim0.3\ \mathrm{M}_\odot$ due to the change in the adiabatic gradient induced by the formation of molecules in the envelope \citep{MS_inflection_0,MS_inflection_1,MS_inflection_2,MS_inflection_3}. This feature is particularly valuable in our fitting process (Section~\ref{sec:4_evaluation}) due to its sensitivity to chemical abundances and dense coverage by our observations. The near infrared isochrone shows a prominent Main Sequence knee at $\sim0.1\ \mathrm{M}_\odot$ where the flux in \texttt{F160W} is suppressed by the onset of $\mathrm{H}_2$ collision-induced absorption \citep{MS_knee_1,MS_knee_2,MS_knee_3}, resulting in bluer colours at lower masses. This overall shift of peak emission towards shorter wavelengths has been spectroscopically observed in L and T subdwarfs \citep{2003ApJ...592.1186B,schneider}. The hydrogen-burning limit (HBL) encompasses another reversal of the colour-magnitude slope in both diagrams at a mass of $\sim0.07\ \mathrm{M}_\odot$ (detailed calculation in Section~\ref{sec:5_predictions} yields $\mathrm{M}_{\mathrm{HBL}}$ = $0.069\ \mathrm{M}_\odot$). As the stellar mass decreases past the limit, the population cools rapidly into the brown dwarf regime. At optical wavelengths, brown dwarfs of lower masses appear marginally bluer immediately after the HBL due to the pressure-broadened K~I line absorption centered at $0.77$~$\micron$ and extending across in the \texttt{F814W} band \citep{2007A&A...465.1085A,2016A&A...589A..21A}.

\begin{figure}[h!]
    \centering
    \includegraphics[width=1\columnwidth]{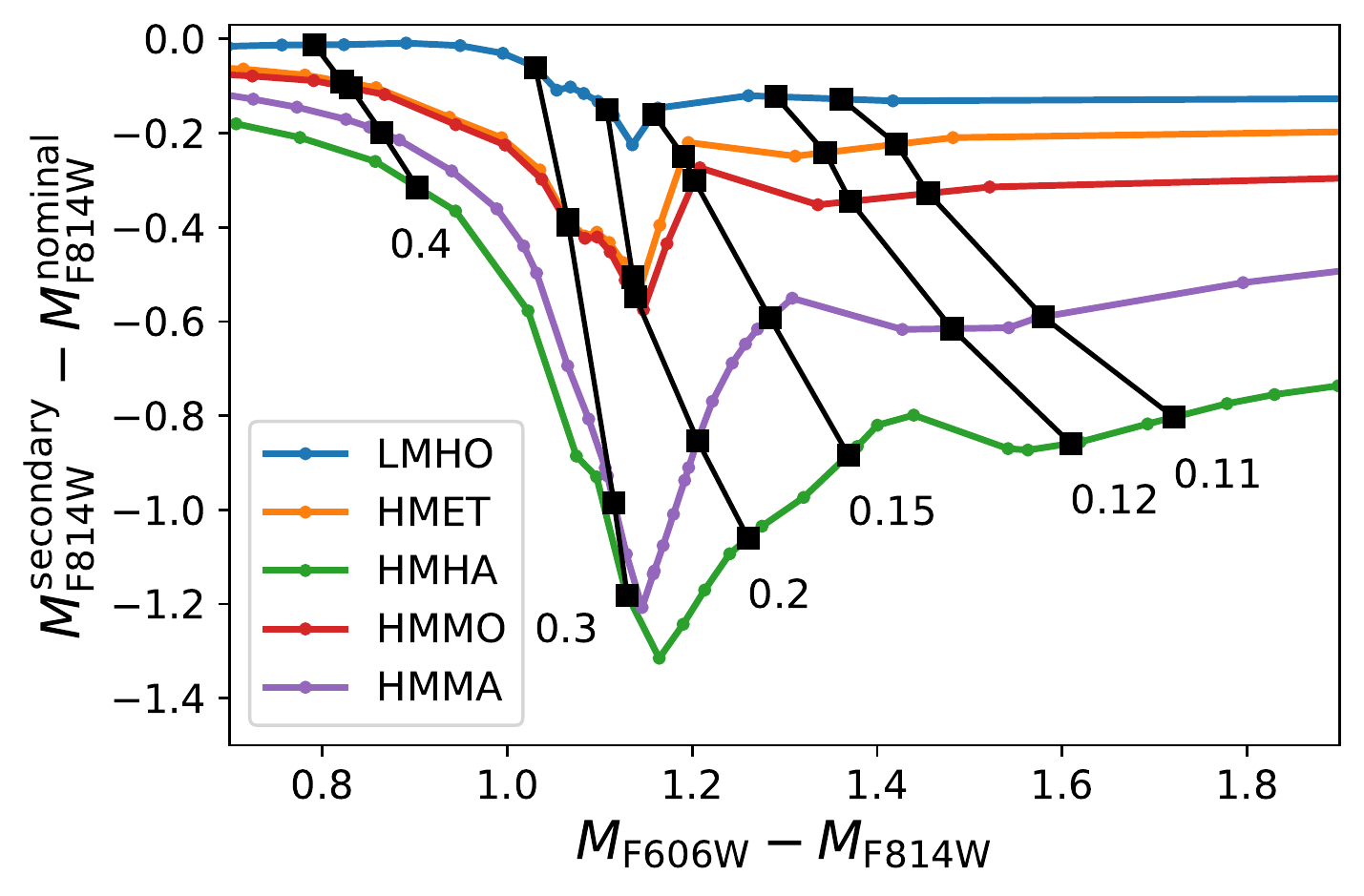}
    \caption{Isochrones for secondary populations listed in Table~\ref{table:table2}. Absolute magnitudes (vertical axis) are displayed as differences after subtracting the absolute magnitude of the nominal population in Fig~\ref{fig:iso_nominal} at the corresponding colour. The range of colours displayed matches the range covered by the available \textit{HST} data, even though some of the isochrones have been calculated at much redder colours. Black lines join the points of equal initial masses along the isochrones that are labeled in solar masses. Extinction effects not included. All magnitudes correspond to \textit{HST} \textit{ACS/WFC} filters.}
    \label{fig:iso_secondary}
\end{figure}

The behavior of secondary populations around the $0.3\ \mathrm{M}_{\odot}$ inflection is shown in Figure~\ref{fig:iso_secondary} as differences to the nominal isochrone in absolute \texttt{F814W} magnitude. In general, all secondary populations are brighter than the nominal one at identical colours, redder at identical masses, and display a more prominent variation in slope. The effect becomes more apparent at higher metallicities and $\alpha$-enhancements, but shows little dependence on the oxygen enhancement alone, suggesting that the lack of a well-defined oxygen peak in Figure~\ref{fig:abundance_hists} is not expected to pose difficulties to isochrone fitting.

\section{Observations} \label{sec:3a_observations}
To determine the best-fit isochrone for \omegacen{}, we
compared each population isochrone to photometric data acquired with \textit{HST} \textit{ACS/WFC} in the \texttt{F606W} and \texttt{F814W} bands (programmes \textit{GO-9444} and \textit{GO-10101}; PI: King), and \textit{HST} \textit{WFC3/IR} in the \texttt{F110W} and \texttt{F160W} bands
(programmes \textit{GO-14118} and \textit{GO-14662} for \textit{WFC3}; PI: Bedin). 
Observations were carried out in a $3'\times3'$ field situated $\sim 3$ half-light radii ($\approx 7'$) southwest of the cluster centre (see field F1 in Figure~1a of \citealt{HST_data}). This is the deepest observed field for \omegacen{} for which both optical and near infrared \textit{HST} observations are available.

\begin{figure}[h!]
    \centering
    \includegraphics[width=1\columnwidth]{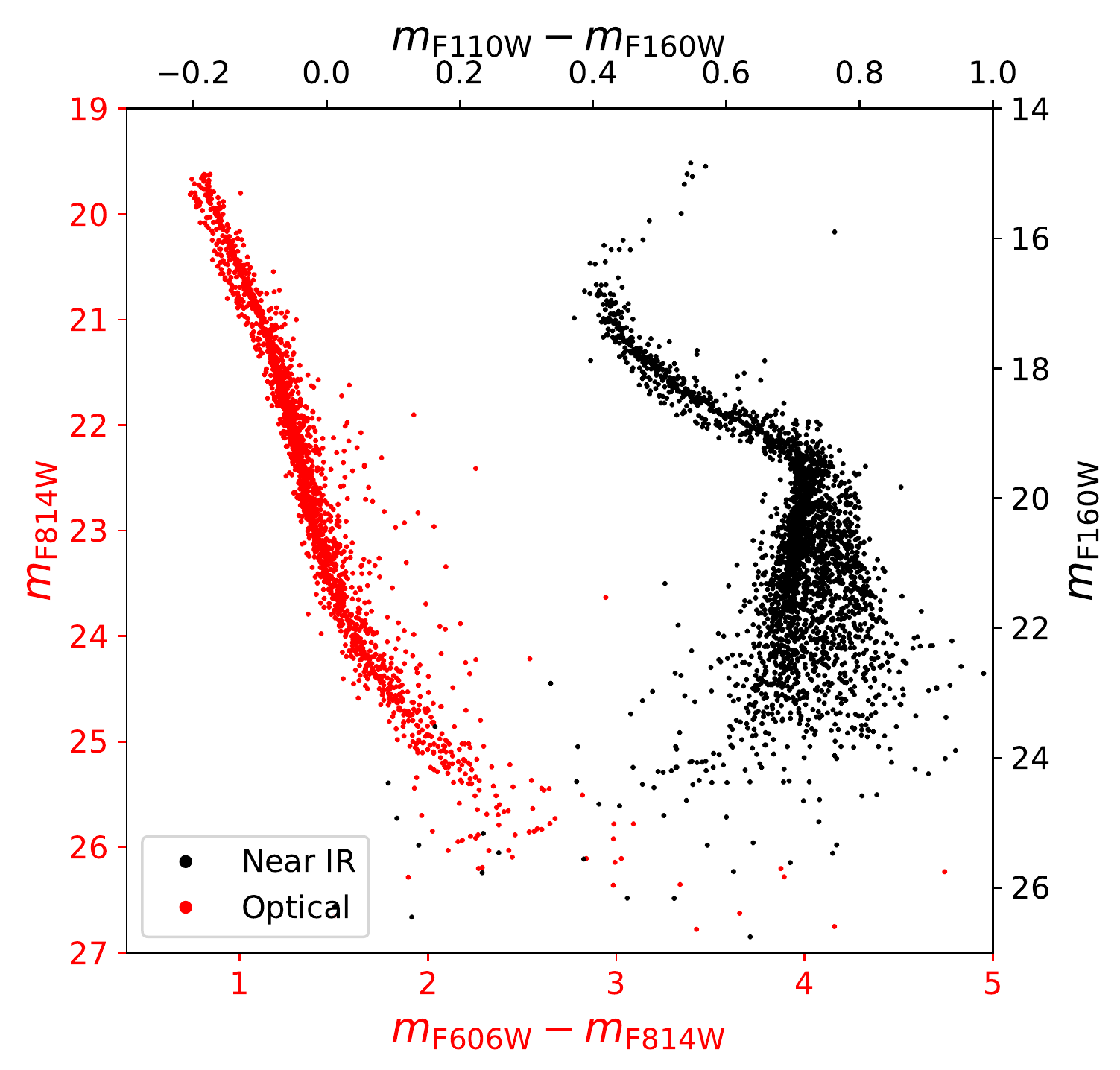}
    \caption{Proper motion-selected zero-pointed differential reddening-corrected photometry of the Main Sequence of \omegacen{}. Optical photometry was acquired with \textit{HST} \textit{ACS/WFC} and near infrared photometry with \textit{HST} \textit{WFC3/IR}. Only unsaturated stars are shown for the optical photometry. The Main Sequence bifurcation can be seen in both datasets.}
    \label{fig:all_data}
\end{figure}

The primary data reduction followed the procedure described in \citet{reduction_3} for two other \textit{HST} \omegacen{} fields, and is analogous to methods adopted in numerous previous works \citep{reduction_1,HST_data,populations,reduction_2,HST_reduction}. 
In brief,
positions, fluxes and multiple diagnostic quality parameters were extracted using the point spread function (PSF) fitting software package \texttt{KS2} \citep{KS2_1,KS2_2}; see \citet{reduction_3} and references therein. The photometric zero-point onto the \texttt{VEGAMAG} system was determined using the approach of \citet{rolly_vegamag}.
The sample was filtered by quality parameters $\sigma$ (photometric error), \texttt{QFIT} (correlation between pixel values and model PSF), and \texttt{RADXS} (flux outside the core in excess of PSF prediction; \citealt{reduction_1,RADXS}),
as described in \citet[Section 4]{reduction_3}.

We used the relative proper motions of sources in the observed region to separate field stars from cluster members. Proper motions were obtained by comparing the extracted positions of stars measured in the earliest and latest  programmes (\textit{GO-9444} and \textit{GO-14662}, respectively), providing epoch baselines of up to $15$ years. Photometry in each filter was corrected for systematic photometric offsets following \citet{rolly_offset}. A general correction for differential reddening was also applied following the method described in \citet[Section 3]{reduction_reddening}.

Measurement of the LF (Section~\ref{sec:4_evaluation}) requires quantification of source completeness as a function of colour and magnitude, for which we followed the approach described in \citet{rolly_offset}. We generated a total of $2.5\times 10^5$ artificial stars (AS) with random positions. For each AS, a \texttt{F606W} magnitude was drawn from a uniform distribution. The remaining three magnitudes (\texttt{F814W}, \texttt{F110W}, \texttt{F160W}) were then chosen to place the AS along the approximate ridgeline of the Main Sequence in various colour-magnitude spaces. AS were introduced in each exposure and measured one at a time to avoid over-crowding, making the process independent of the LF. A star was considered recovered when the difference between the generated and measured star position was less than $0.1$ pixels and the magnitude difference was less than $0.4\ \mathrm{mag}$. Finally, the stars were divided into half-magnitude bins and the photometric errors and completeness for each bin were computed.

The near infrared and optical colour-magnitude diagrams based on our observations are shown in Figure~\ref{fig:all_data}. The full catalogue of source astrometry, photometry, membership, and completeness is provided as an associated data product and described more fully in Appendix~\ref{sec:7b_catalog}.

\section{Evaluation} \label{sec:4_evaluation}
With mass-luminosity and colour-magnitude sequences computed for multiple populations, we were able to determine the optimal isochrone and IMF by comparing the predictions of those models to the \textit{HST}-observed Main Sequence at optical and near infrared wavelengths.

\subsection{Best-fit isochrone}

\begin{figure}[h!]
    \centering
    \includegraphics[width=1\columnwidth]{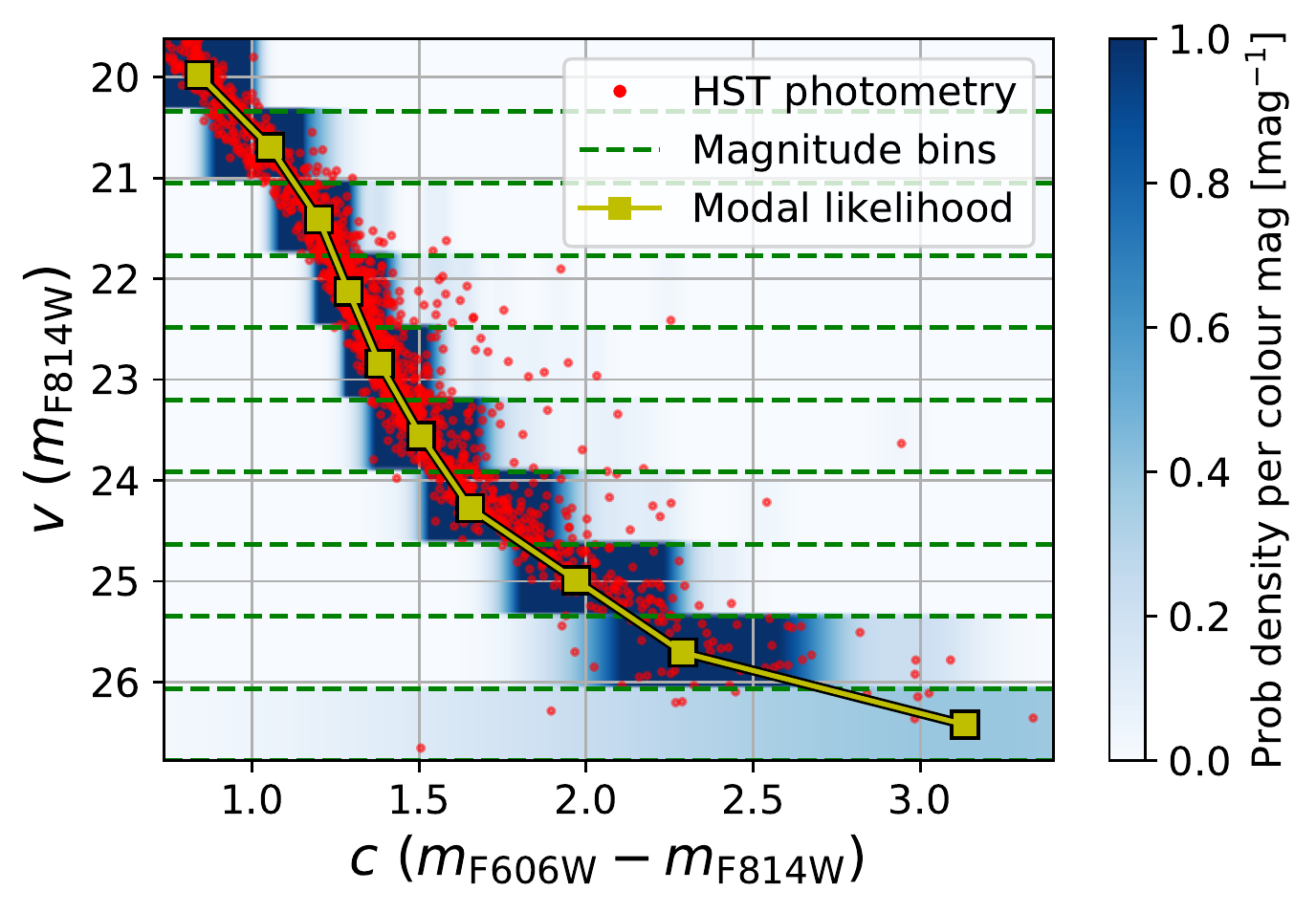}
    \includegraphics[width=1\columnwidth]{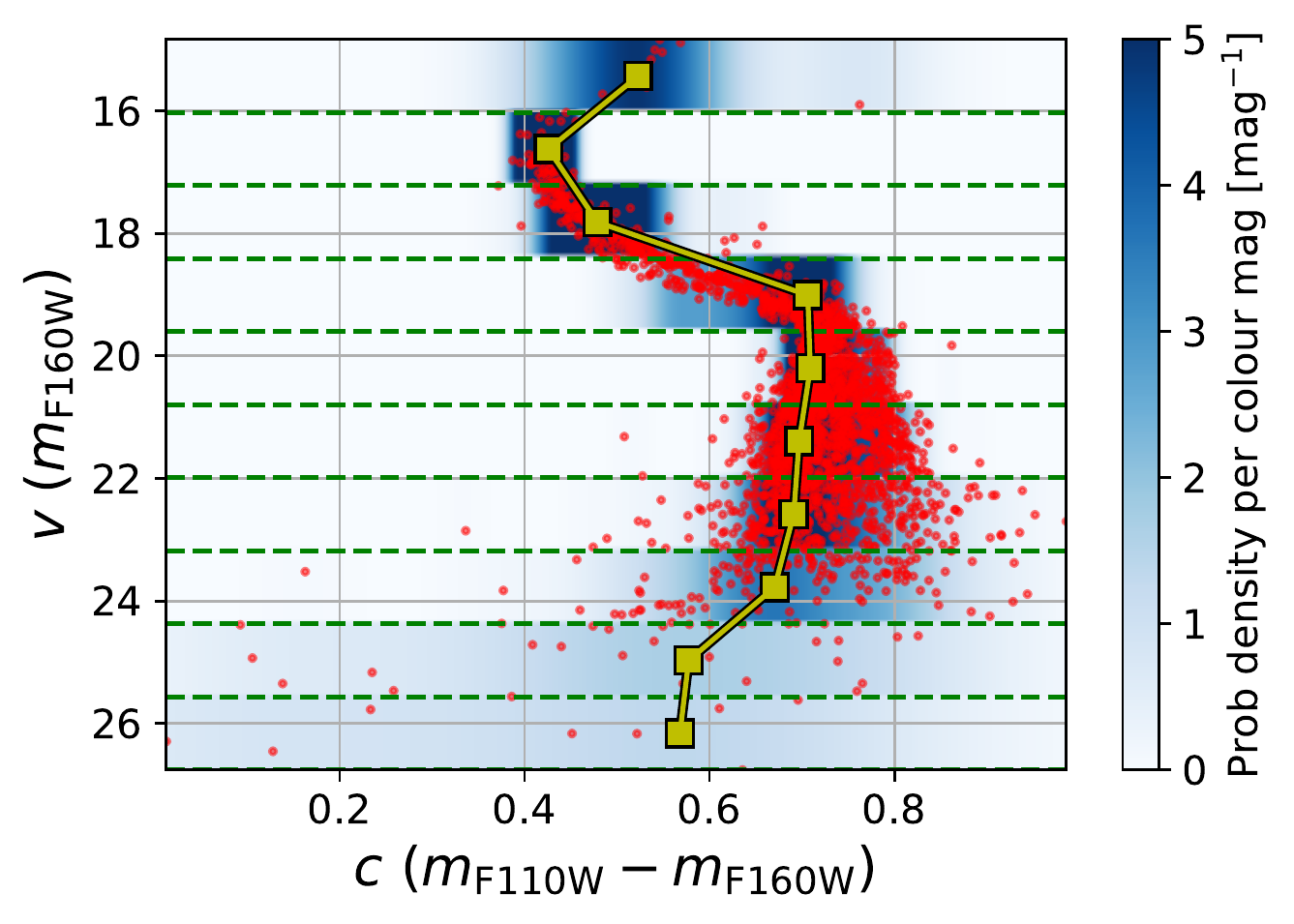}
    \caption{Colour probability distributions inferred from the observed scatter in \textit{HST} photometry. Also shown are the boundaries of the magnitude bins used in our fitting analysis. Yellow markers indicate the mode of the distribution in each bin. \textit{Top}: Optical data from \textit{HST} \textit{ACS/WFC}. \textit{Bottom}: Near infrared data from \textit{HST} \textit{WFC3/IR}.}
    \label{fig:gaussian}
\end{figure}


We adopted a distance modulus of $13.60\pm0.05$ based on the distance to \omegacen{} of $5.24\pm0.11\ \mathrm{kpc}$ derived by \citet{omega_parallax} from the parallaxes of $\sim7\times10^4$ members. The adopted value is marginally smaller than the distance modulus of $13.69$ derived by \citet{distance_modulus} from isochrone fit to the CMD.

We sought an isochrone that is most statistically compatible with the observed photometry, accounting for the average spread in the data introduced by unmodelled astrophysical and instrumental phenomena, such as the variation in abundances across the cluster, multiple star systems, observational errors, etc. First, we developed a likelihood model that predicts the probability of finding a cluster member at a given point $(c,v)$ in colour-magnitude space assuming that the average population is well-described by one of our isochrones:

\begin{equation}
\begin{split}
\mathcal{P}_E(c,v)\propto \int{\xi(m)P\left(c,v \Bigm| c_0(m,E), v_0(m,E)\right) dm}
\end{split}
\label{eq6}
\end{equation}

\noindent In the equation, $P(...)$ is the probability of observing a member at $(c,v)$ assuming that the ``true'' location of the star in the colour-magnitude space (including reddening) is $(c_0,v_0)$. Both $c_0$ and $v_0$ are functions of the initial stellar mass, $m$, and the optical interstellar reddening $E$. Finally, $\xi(m)$ is the IMF, such that $\xi(m)dm$ is the number of stars in the cluster with masses between $m$ and $m+dm$. Note that a proportionality sign is used here as the likelihood function is not appropriately normalized in the given form.

\begin{figure}[h!]
    \centering
    \includegraphics[width=1\columnwidth]{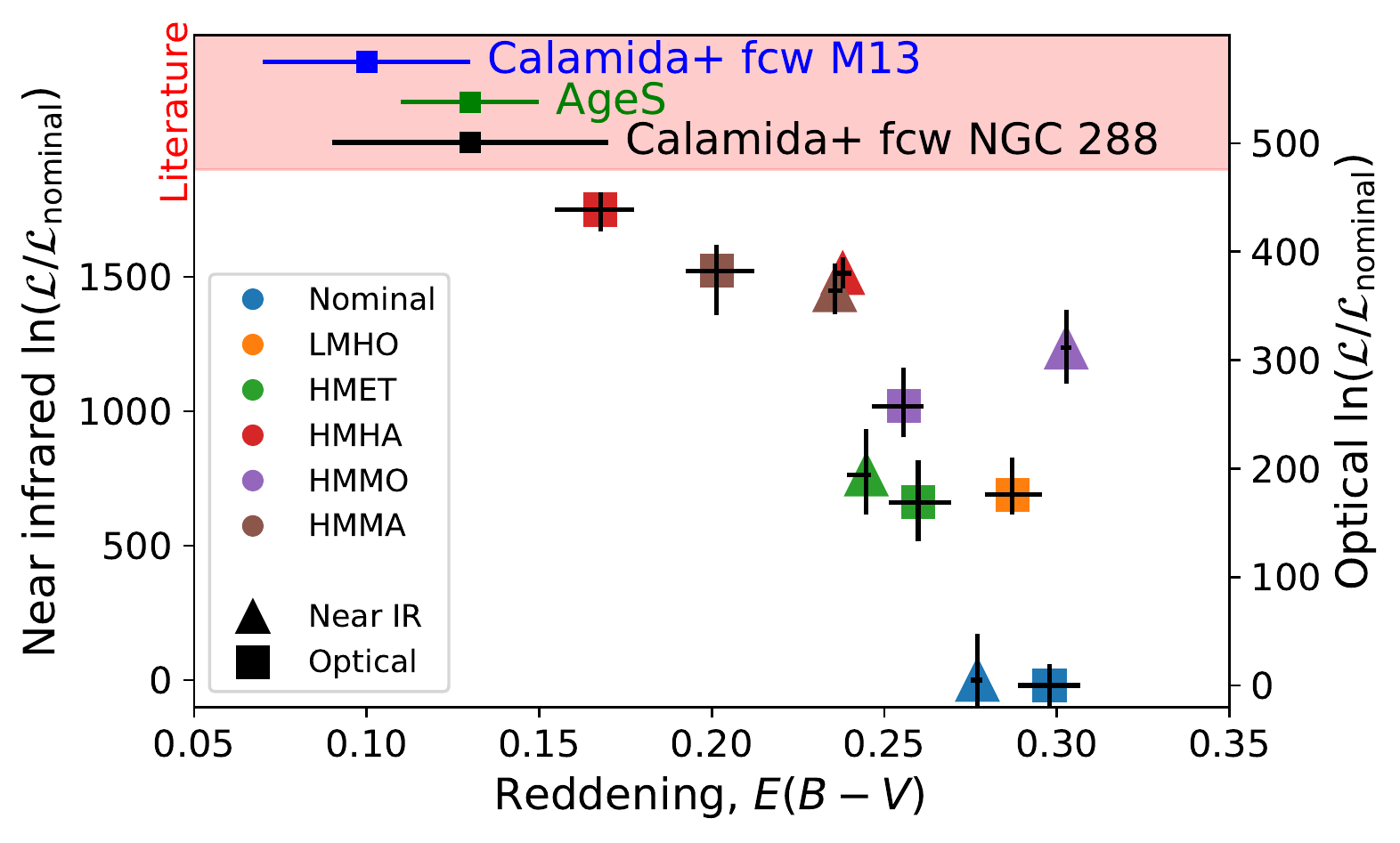}
    \caption{Likelihoods of compatibility and best-fit interstellar reddening for the population isochrones calculated in this study based on \textit{HST} photometry. ``Nominal'' refers to the nominal population described in Section~\ref{sec:3_isochrones}. Secondary populations are summarized in Table~\ref{table:table2}. The vertical axes are normalized to $\mathcal{L}_\mathrm{nominal}$. Error bars indicate random errors in the values as described in text. Selected reddening values from literature are also shown with their uncertainties. \textit{AgeS} refers to the reddening estimate used in \citet[][green]{AgeS}. The other two values are taken from \citet{omega_cen_reddening}, calculated \textit{from comparison with} (fcw) NGC 288 (black) and M13 (blue). Marker shapes differentiate between fits obtained using optical \textit{ACS/WFC3} (squares; right axis) and near infrared \textit{WFC3/IR} photometry (triangles; left axis). The two sets of values have different vertical scaling and cannot be compared against each other.}
    \label{fig:reddening}
\end{figure}

\begin{figure}[h!]
    \centering
    \includegraphics[width=1\columnwidth]{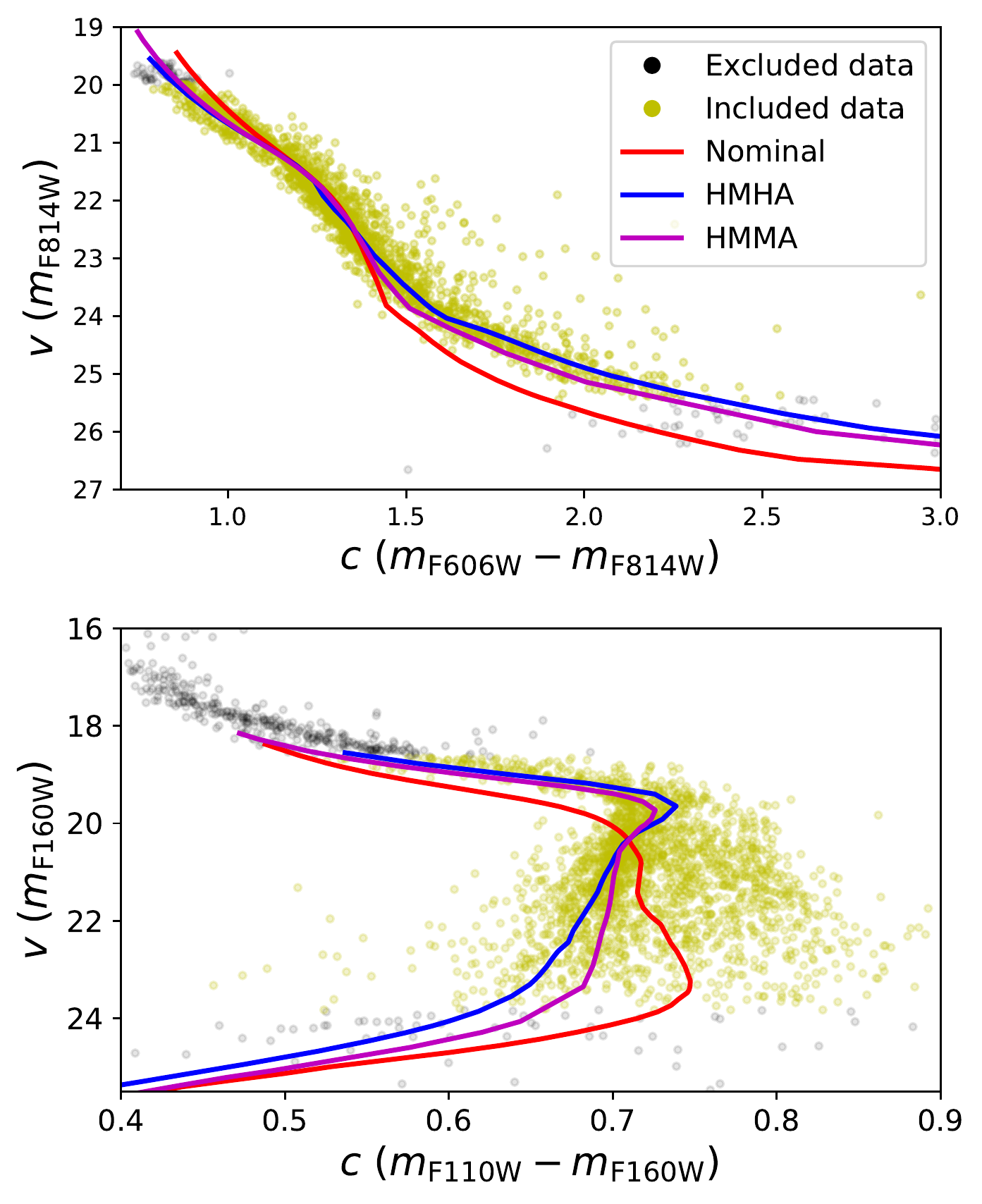}
    \caption{
    Nominal (Table~\ref{table:fixed_properties}) and two secondary (Table~\ref{table:table2}) population isochrones overplotted on \textit{HST} photometry. The isochrones have been adjusted by the best-fit reddening values. The colour of markers indicates whether any particular member was or was not included in the log-likelihood optimization described in text to evaluate the accuracy of the isochrone. \textit{Top}: Optical data from \textit{HST} \textit{ACS/WFC}. \textit{Bottom}: Near infrared data from \textit{HST} \textit{WFC3/IR}.}
    \label{fig:iso_vs_data}
\end{figure}

The individual probability distribution, $P(...)$, encapsulates the scatter of photometry around the best-fit isochrone, and must account for all relevant effects including experimental uncertainties, unresolved multiple stars, and multiple distinct populations known to be present in \omegacen{}. For our purposes, both $P(...)$ and $\xi(m)$ can be estimated empirically from the observed spread of \textit{HST} data across the colour-magnitude space without theoretical input. In this method, the scatter along the colour axis is degenerate with that along the magnitude axis, as any observed distribution of data points may be reproduced by perturbing predicted photometry along only one axis and not the other. We therefore chose to sample the observed scatter in photometry along the colour axis only and use the magnitude axis as an estimator of the initial stellar mass by interpolating the theoretical mass-luminosity relation for the population under evaluation.

The empirical scatter was sampled from the observed data as follows. First, the range of apparent magnitudes in $v$ (\texttt{F814} in the optical, \texttt{F160W} in the near infrared) was divided into $10$ bins of equal widths as demonstrated in Figure~\ref{fig:gaussian}. Within each bin, the variation of magnitude was ignored and the probability density function (PDF) of the colour distribution was computed using Gaussian kernel density estimation with bandwidths calculated as in \citet{gauss_kde}. The distribution was then translated along the colour axis to place the mode at the origin. The inferred PDF around the mode was then used as the scatter in colour for all stars whose magnitudes fall within the magnitude bin.

The initial mass function in equation (\ref{eq6}), $\xi(m)$, was evaluated by converting all measured magnitudes in the \textit{HST} dataset to initial stellar masses using the linearly interpolated mass-magnitude relations derived from stellar models discussed in Section~\ref{sec:3_isochrones}. The inferred distribution of masses was then converted into the mass PDF, $\xi(m)$, using the same kernel density estimation method as in the colour spread \citep{gauss_kde}, but trimmed on both sides at the lowest and highest modelled stellar masses respectively to avoid extrapolation.

The integral in equation (\ref{eq6}) was computed numerically by drawing $10^4$ masses from the inferred $\xi(m)$ PDF, evaluating the integrand for each and summing the results. Finally, the total likelihood of a given isochrone being compatible with the \textit{HST} dataset ($\mathcal{L}(E)$) was calculated as in equation (\ref{eq7}).

\begin{equation}
\begin{split}
\mathcal{L}(E)\propto \prod_i\mathcal{P}_E(c_i,v_i)
\end{split}
\label{eq7}
\end{equation}

\noindent In the equation, the product may, in principle, be taken over all individual measurements $(c_i,v_i)$. In practice, we must only include those members in the \textit{HST} dataset that fall within the magnitude range of \textit{all} calculated isochrones, as stellar masses of members out of range cannot be reliably estimated. Furthermore, since inferred stellar masses are dependent on interstellar reddening which is not \textit{a priori} known, we must only select those cluster members for analysis that fall within the modelled range at all realistic reddenings, which we conservatively take to be $E(B-V)\in[0.0, 0.4]$. Our final choice of bounds was $v\in(18.65,23.83)$ in the near infrared (\textit{WFC3/IR} \texttt{F160W}) and $v\in(19.99,25.44)$ in the optical (\textit{ACS/WFC} \texttt{F814W}), accommodating approximately $85\%$ and $94\%$ of all available measurements, respectively. The subset of selected members is shown in Figure~\ref{fig:iso_vs_data}.

For the nominal and each of the secondary populations, we maximize $\mathcal{L}(E)$ with respect to the interstellar reddening, $E(B-V)$. We estimate the random error in the best-fit reddening value, $E_0$ by considering three contributions. The intrinsic fitting error may be adopted as the Cramér–Rao bound:
\begin{equation}
\mathrm{Var}\left(E_0\right)=-\left( \frac{\delta^2 \ln\mathcal{L}}{\delta E^2} \Bigm|_{E=E_0} \right)^{-1}
\label{eq9}
\end{equation}
which in our case evaluated to $\sqrt{\mathrm{Var}\left(E_0\right)}\approx0.001$ for all isochrones. The contributions of the random sampling of $\xi(m)$ during numerical integration and experimental uncertainties in the data were estimated by repeating the fitting process $10$ times with different samples of $\xi(m)$ and random Gaussian perturbations in the data. Finally, the error induced by the uncertainty in the distance to the cluster was determined by repeating the fitting process for upper and lower $1\sigma$ bounds on the distance modulus value.

All of the aforementioned contributions were combined in quadrature. The resulting likelihoods and best-fit reddening values are shown in Figure~\ref{fig:reddening} with uncertainties. Every secondary isochrone performs better than the nominal one, with \texttt{HMHA} offering the best fit in both optical and near infrared wavelengths. As such, we used \texttt{HMHA} for our predictions of brown dwarf photometry described in Section~\ref{sec:5_predictions}. The best-fit reddening values corresponding to this isochrone are $E(B-V)=0.238\pm0.003$ from the near infrared data and $E(B-V)=0.17\pm0.01$ from the optical data. The random errors in both $E(B-V)$ estimates quoted here and shown in Figure~\ref{fig:reddening} are likely not representative of the true uncertainty in the value, which is primarily driven by systematic effects due to the simplified population parameters, the reddening law, and errors intrinsic to the calculated stellar models. The scatter in $E(B-V)$ estimates between the optical and near infrared datasets suggests that the true value of the uncertainty in reddening is of the order of $\sim 0.07$.

Our reddening estimates exceed most literature values, of which three are shown in Figure~\ref{fig:reddening}. The \textit{Cluster AgeS} experiment \citep{AgeS} uses the value of $E(B-V)=0.13\pm0.02$ based on the value of $E(B-V)=0.132$ given by the map of dust emission from \citet{dust_map} at a particular point within \omegacen{} and assuming the uncertainty of $0.02$ motivated by the variation of reddening across the cluster. In \citet{omega_cen_reddening}, two reddening values of $E(B-V)=0.13\pm0.04$ and $E(B-V)=0.10\pm0.03$ are derived from comparison with NGC 288 and M13, respectively. The apparent discrepancy in reddening values may be an artifact of our approach, since a single population is used to model both Main Sequences of the cluster. Consequently, the large helium fraction adopted in this study makes the predicted colours around the Main Sequence knee bluer, corresponding to a higher best-fit reddening value.

Two best-fitting secondary isochrones -- \texttt{HMHA} and \texttt{HMMA} -- as well as the nominal isochrone are plotted against the \textit{HST} data in Figure~\ref{fig:iso_vs_data}, visually illustrating the goodness-of-fit. Both isochrones in the figure have been corrected for the corresponding best-fit interstellar reddening parameters.

\subsection{Best-fit LF and IMF}

\begin{figure}[h!]
    \centering
    \includegraphics[width=1\columnwidth]{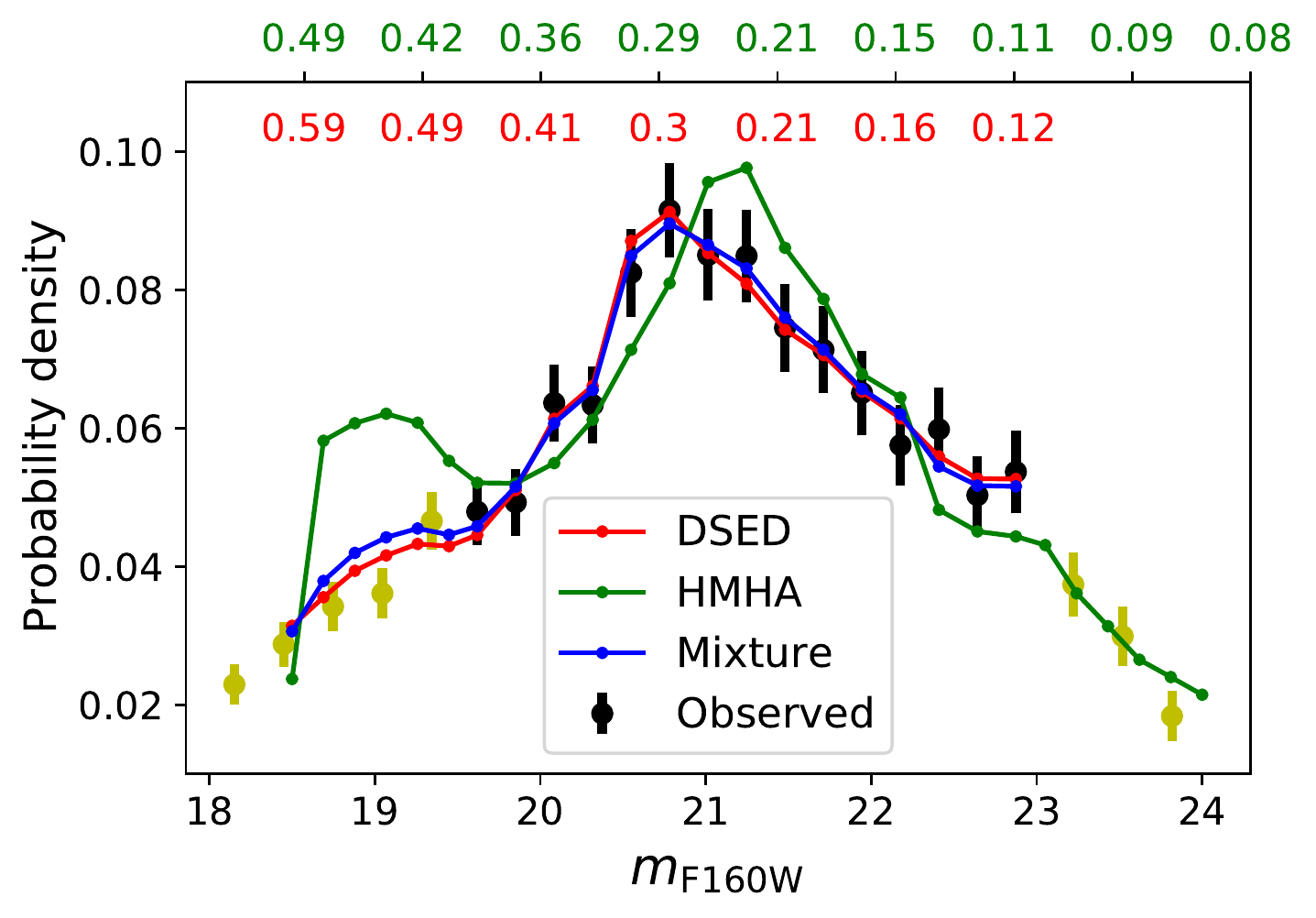}
    \caption{Observed luminosity function (LF) for \omegacen{} (black) with three theoretical fits corresponding to the cases of $\mu=0$ (red, solar helium population only), $\mu=1$ (green, enhanced helium population only) and $\mu$ being a free parameter (both populations). The enhanced helium population is based on the mass-luminosity relation of the best-fit isochrone calculated in this study, \texttt{HMHA}. The solar helium population is based on the mass-luminosity relationship from \citet[][\textit{DSED}]{DSED}. In each case, a broken power law IMF is assumed (Equation \ref{eq:powerlaw}) with the best-fit values $\gamma=0.89\pm0.06$, $\gamma=0.50\pm0.07$ and $\gamma=0.83\pm0.08$ for the three cases respectively. The best-fit mixing fraction in the case of two populations was calculated as $\mu=0.15\pm0.14$. The fitting is carried out between the apparent magnitudes of $19.5$ and $23$ only since \textit{DSED} models are not available at the faint end and photometry becomes increasingly unreliable at the bright end due to saturation \citep{reduction_3}. Nonetheless, the observed LF outside this range is shown in yellow for completeness. The normalization on the vertical axis is such that the sum of all bins used in the fit is unity. The upper colour-coded horizontal axis indicates the initial stellar masses corresponding to magnitudes for the solar helium population (red) and the enhanced helium population (green) in solar masses.}
    \label{fig:LF}
\end{figure}

Assuming the best-fit (\texttt{HMHA}) isochrone to be representative of the average distribution of \omegacen{} members in colour-magnitude space, we now seek a suitable IMF for the cluster to estimate the population density. As will be demonstrated shortly, the cluster is well described by a broken power law:

\begin{equation}
    \xi(m)\propto\begin{cases}
        m^{-2.3},& \text{if } m>0.5\ \mathrm{M}_\odot\\
        m^{-\gamma},& \text{if } m\leq0.5\ \mathrm{M}_\odot\\
    \end{cases}
    \label{eq:powerlaw}
\end{equation}

The power index of the high-mass regime ($-2.3$) as well as the breaking point ($m=0.5\ \mathrm{M}_\odot$) are fixed to the values employed in the ``universal'' IMF derived in \citet{Kroupa}. It has been demonstrated by \citet{omega_cen_IMF} that those values are well-suited to the high-mass regime of \omegacen{}. The power index of the low-mass regime ($-\gamma$) is allowed to vary. For comparison, \citet{omega_cen_IMF} use $\gamma=0.8$, while the ``universal'' IMF introduces additional breaking points with different power indices.

The theoretical mass-luminosity relationship for \texttt{HMHA} was combined with the IMF to derive the theoretical luminosity function (LF) for \omegacen{} as a function of $\gamma$. The best value of $\gamma$ was determined by optimizing the $\chi^2$ statistic for the goodness-of-fit between the theoretical and observed LFs. Our analysis of the LF was carried out in the \texttt{F160W} band of \textit{HST} \textit{WFC3/IR} between the apparent magnitudes of $19.5$ and $23$. Within this range, the data were divided into $15$ uniform bins with the count uncertainty in each bin taken as the square root of the count. The counts have also been adjusted for estimated sample completeness in each bin as discussed in Section~\ref{sec:3a_observations}. The histogram was normalized and used as an estimate of the underlying PDF.

The theoretical LF was calculated from the IMF in equation (\ref{eq:powerlaw}) using the mass-luminosity relationship from \texttt{HMHA} and integrating the resulting PDF within each magnitude bin. Both observed and theoretical LFs within the fitting range are plotted in Figure~\ref{fig:LF} in green and black respectively for the best-fit value of $\gamma=0.50\pm0.07$. The correspondence between the two LFs appears poor, indicating that the \texttt{HMHA} population alone cannot reproduce the observed LF. This result is not surprising as our best-fit isochrone was calculated for the helium mass fraction of the blue sequence in \omegacen{} that is only representative of a minority of the members.

To improve the fit, we added a second population with a solar helium mass fraction and a mass-luminosity relationship adopted from the Dartmouth Stellar Evolution Database (\textit{DSED}; \citealt{DSED})  for $[\mathrm{M}/\mathrm{H}]=-1.7$ and $\mathrm{Y}=0.2456$. The extinction of $0.085\ \mathrm{mag}$ was applied to synthetic \texttt{F160W} photometry from \textit{DSED} based on the average magnitude difference between the best-fit reddening ($E(B-V)=0.17$, lower bound most consistent with literature) and reddening-free ($E(B-V)=0$) \texttt{HMHA} isochrones. The mixing fraction between the two populations, $\mu$, was treated as a free parameter varying between $0$ (\textit{DSED} population only) and $1$ (\texttt{HMHA} only). The best-fit LF based on both \texttt{HMHA} and \textit{DSED} as well as the best-fit based on \textit{DSED} alone ($\mu=0$) are shown in Figure~\ref{fig:LF}. The calculated best-fit value of $\mu=0.15$, is comparable to its uncertainty of $\pm0.14$. Therefore, we present this result as the $2\sigma$ upper limit on the blue sequence population fraction, $\mu<0.45$. The blue sequence thus contributes less than $45\%$ of the cluster population in the observed region, in agreement with \citet{discrete_populations_experimental}. The best-fit value of $\gamma$ when both \texttt{HMHA} and \textit{DSED} LFs are included is $0.83\pm0.08$, which matches the adopted value of $\gamma$ in \citet{omega_cen_IMF}.

\section{Predictions} \label{sec:5_predictions}
\subsection{Substellar population of \omegacen{}}

\begin{figure*}
    \centering
    \includegraphics[width=2\columnwidth]{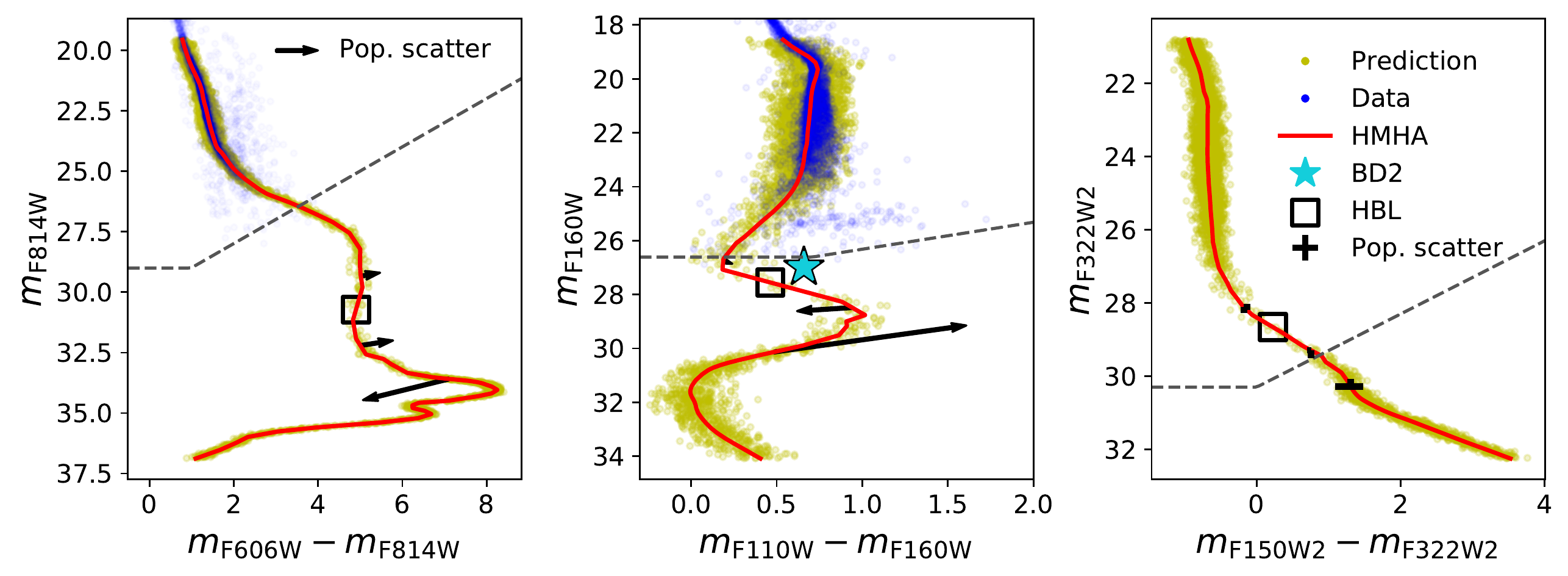}
    \caption{Predicted colour-magnitude diagrams for \omegacen{} from the Main Sequence through the stellar/substellar gap and down to the appearance of first brown dwarfs. Predicted points are based on the best-fit \texttt{HMHA} isochrone and the best-fit IMF. For clarity, a Gaussian scatter in colour by $0.1$ magnitudes was added to each point. Where available, predicted CMDs are shown alongside existing \textit{HST} photometry reaching the cool end of the Main Sequence. All CMDs are normalized to $1700$ objects between $0.1\ \mathrm{M}_\odot$ and $0.3\ \mathrm{M}_\odot$. The instruments used are \textit{HST} \textit{ACS/WFC} (\textit{left}), \textit{HST} \textit{WFC3/IR} (\textit{middle}) and \textit{JWST} \textit{NIRCam} (\textit{right}). The cyan star shows the near infrared colour and magnitude of BD2 -- a candidate brown dwarf in the globular cluster M4 from \citet{BD_hunt_2}. The magnitude of BD2 shown here has been adjusted for the difference between the distance moduli of \omegacen{} and M4 using the distance measurement from \citet{M4_distance}. The grey dashed line indicates the approximate faint limit for both \textit{HST} datasets and the expected faint limit of future \textit{JWST} measurements, calculated using the \textit{JWST} Exposure Time Calculator \citep{JWST_ETC} for a $1\ \mathrm{hr}$ exposure and signal-to-noise ratio of $2$. The colour and magnitude corresponding to the hydrogen-burning limit (HBL) are highlighted in each case. The arrows in the \textit{HST} plots (\textit{left} and \textit{middle}) indicate the approximate direction and relative magnitude of the effect of decreasing metallicity and $\alpha$-enhancement as estimated from the difference between the best-fitting \texttt{HMHA} and \texttt{HMMA} secondary populations as well as the nominal population. In the case of \textit{JWST}, the scatter among isochrones does not display a clear direction and is shown with error bars instead.}
    \label{fig:prediction_CMD}
\end{figure*}

\begin{figure}[h!]
    \centering
    \includegraphics[width=1\columnwidth]{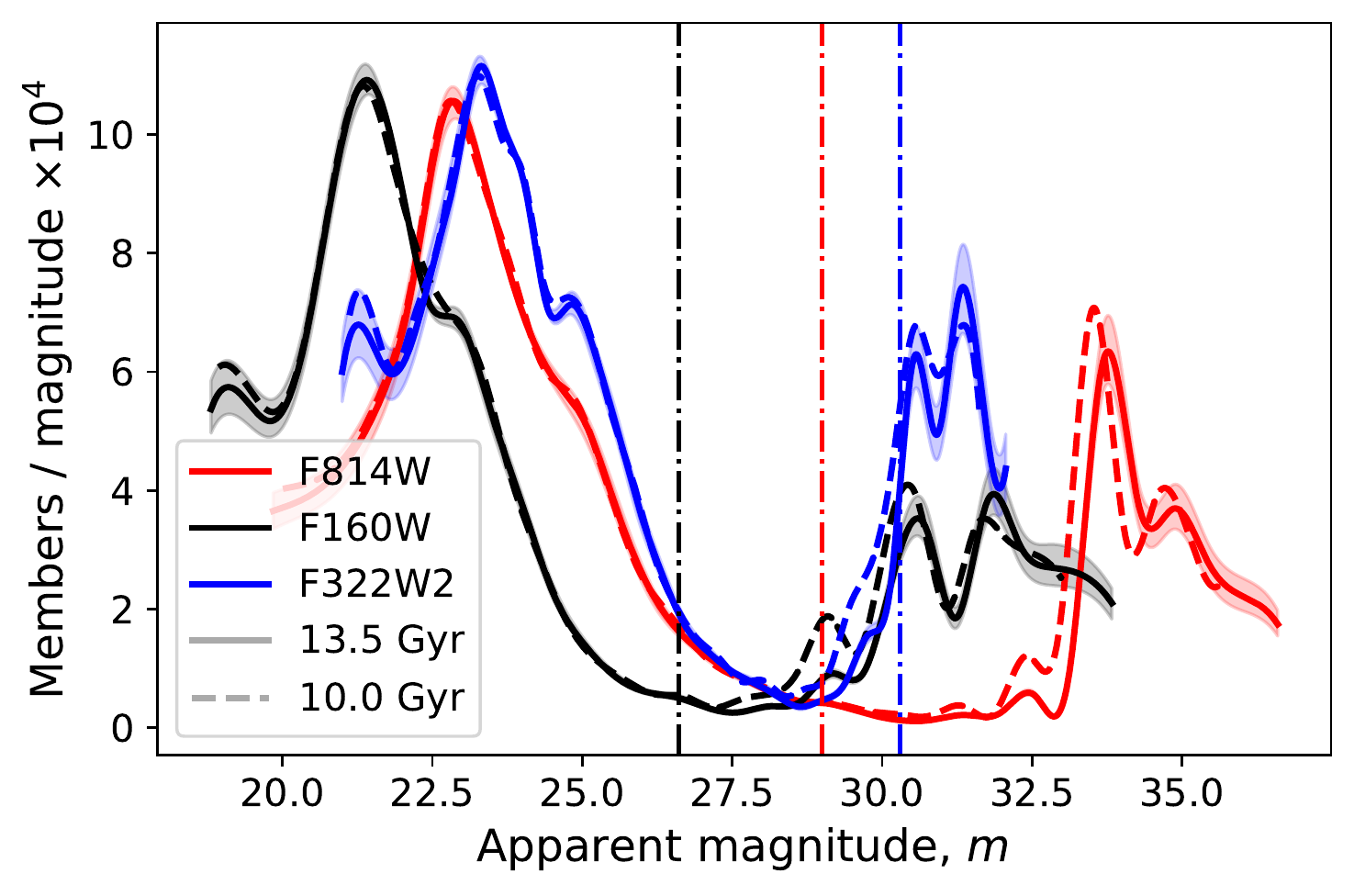}
    \caption{Predicted luminosity function for the \textit{HST} \textit{ACS/WFC} \texttt{F814W} band, the \textit{HST} \textit{WFC3/IR} \texttt{F160W} band, and the \textit{JWST} \textit{NIRCam} filter \texttt{F322W2}. Curves are shown for the population ages of $13.5\ \mathrm{Gyr}$ (solid) and $10\ \mathrm{Gyr}$ (dashed). The two peaks correspond to the Main Sequence and brown dwarfs in the cluster with a stellar/substellar gap in between. 
    The vertical dash-dotted lines indicate the approximate faint limits for the instruments shown, calculated identically to Figure~\ref{fig:prediction_CMD}. The shaded areas around the curves indicate the range of our predictions based on the uncertainty in the determined IMF. While the shown ranges are for the age of $13.5\ \mathrm{Gyr}$, similar uncertainties apply to the case of $10\ \mathrm{Gyr}$. This figure demonstrates the superiority of infrared observations with \textit{JWST} as the apparent magnitude of brown dwarfs enters the limit of the instrument.}
    \label{fig:prediction_LF}
\end{figure}

\begin{figure}[h!]
    \centering
    \includegraphics[width=1\columnwidth]{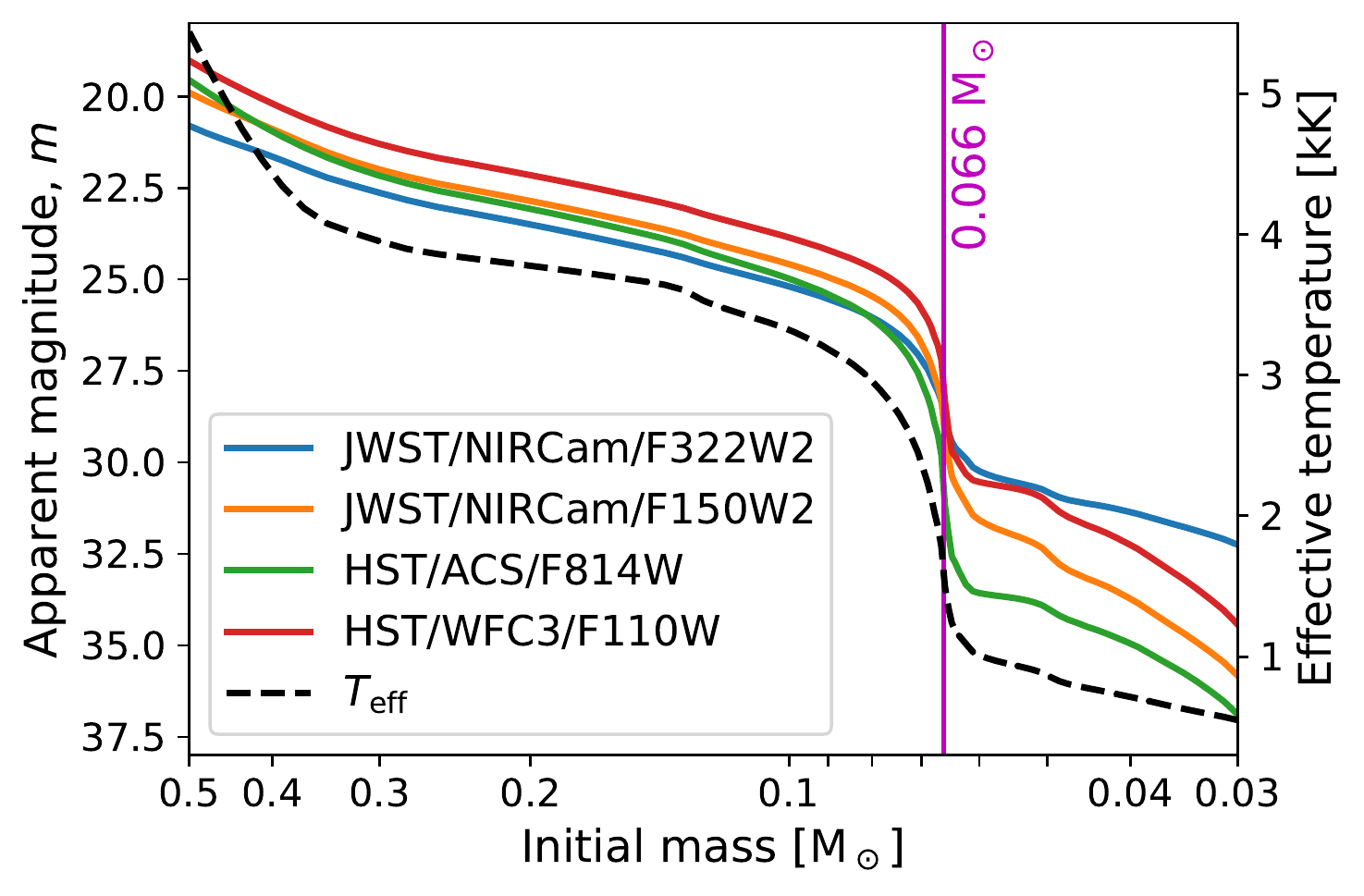}
    \caption{
    Predicted mass-luminosity relations for the same set of instruments as in Figure~\ref{fig:prediction_LF} as well as the predicted mass-effective temperature relationship (dashed). The approximate mass of the hydrogen-burning limit is highlighted with a vertical line and labelled.}
    \label{fig:prediction_ML}
\end{figure}

In this section, we present our predictions of colours, magnitudes and CMD densities of brown dwarfs in \omegacen{} using the best-fit isochrone (\texttt{HMHA}) and the best-fit IMF (Equation \ref{eq:powerlaw}) calculated in Section~\ref{sec:4_evaluation}. Figure~\ref{fig:prediction_CMD} shows predicted CMDs for the cluster in three different sets of filters: \texttt{F814W} vs \texttt{F606W}-\texttt{F814W} for \textit{HST} \textit{ACS/WFC}, \texttt{F160W} vs \texttt{F110W}-\texttt{F160W} for \textit{HST} \textit{WFC3/IR} and \texttt{F322W2} vs \texttt{F150W2}-\texttt{F322W2} for \textit{JWST} \textit{NIRCam}.

For the first two diagrams, observed Main Sequence photometry is available and shown alongside predicted colours and magnitudes in blue. The density of points in the predicted set is proportional to the PDF luminosity function (Figure~\ref{fig:LF}) extended into the brown dwarf regime. The normalization is such that approximately $1700$ points fall between the initial masses of $0.1\ \mathrm{M}_\odot$ and $0.3\ \mathrm{M}_\odot$. This choice closely matches the number of members within the same range of masses in the optical \textit{HST} dataset used in this analysis. For clarity, a Gaussian spread with a standard deviation of $0.1$ magnitudes was applied to each point from the predicted set along the colour axis. Each CMD contains a region of low source density below the cool end of the Main Sequence (the stellar/substellar gap) followed by an increase in density at fainter magnitudes, corresponding to the accumulation of cooling brown dwarfs.

The effects of metallicity and $\alpha$-enhancement on the predicted brown dwarf colours and magnitudes are indicated by error bars and arrows in Figure~\ref{fig:prediction_CMD} at three effective temperatures: $1900\ \mathrm{K}$, $1300\ \mathrm{K}$, and $1000\ \mathrm{K}$. The first temperature is just above the HBL, while and the latter two are below the HBL. 
The effect size was inferred from the scatter among the two best-fit isochrones, \texttt{HMHA} and \texttt{HMMA}, and the nominal isochrone. Note that at $T_\mathrm{eff}=1000\ \mathrm{K}$, only the best-fit isochrone (\texttt{HMHA}) has computed atmosphere models, so scatter at this temperature is 
based on extrapolated bolometric corrections for both \texttt{HMMA} and the nominal isochrones and may be unreliable. The scatter is substantial in the optical and near infrared \textit{HST} bands, 
but appears far less significant in infrared \textit{JWST} bands,
as the \texttt{F150W2} band accommodates most of the prominent metallicity features in the spectrum (see Figure~\ref{fig:example_spectra}).
A different choice of narrow-band filters would make colour measurements more sensitive to chemical abundances at the expense of worse signal-to-noise ratio.

The extended luminosity functions that the CMD predictions are based on are shown in Figure~\ref{fig:prediction_LF}. As before, both the main peak corresponding to the Main Sequence and the brown dwarf peak just emerging at the faint end can be seen with a gap in between. In the figure,
%
each plot is given for two cluster ages of $10\ \mathrm{Gyr}$ and $13.5\ \mathrm{Gyr}$ corresponding to the expected ages of the youngest and oldest members in \omegacen{}. While the Main Sequence peaks appear relatively unaffected by age, the brown dwarf peaks emerge at slightly brighter magnitudes at $10\ \mathrm{Gyr}$. As expected for objects in energy equilibrium, the Main Sequence evolves slowly with time. By contrast, substellar objects have entered their cooling curves and are moving steadily across colour-magnitude space. 
Hence, the luminosity function gap for \omegacen{} and other globular clusters provides 
a potential age diagnostic for the system, assuming the evolutionary timescales are correctly modeled
\citep{age_gap,whitepaper,2004ApJS..155..191B,2009IAUS..258..317B}. We also show in Figure~\ref{fig:prediction_LF} the variance in LF predictions taking into account uncertainty in the inferred IMF.  In general, a higher value of the power index results in fewer low-mass members in the cluster and vice versa. Note that the width of the stellar/substellar gap is insensitive to the adopted IMF and is primarily determined by the mass-effective temperature relationship of the population.
The normalization in Figure~\ref{fig:prediction_LF} is for the \textit{total} number of helium-enriched members in the entire cluster based on the best-fit IMF (Equation \ref{eq:powerlaw}), the best-fix mixing ratio ($\mu=0.15$) and the assumed total cluster mass of $4\times10^6\ \mathrm{M}_\odot$ \citep{omega_cen_total_mass}.

Finally, we provide a set of mass-luminosity relations for the aforementioned \textit{JWST} and \textit{HST} filters in Figure~\ref{fig:prediction_ML} alongside the mass-effective temperature relationship. All curves are based on the best-fit isochrone (\texttt{HMHA}). The predicted initial stellar mass at the hydrogen-burning limit (HBL) 
was taken as the mass for which the total proton-proton chain luminosity output corresponds to a half of the total luminosity output at $13.5\ \mathrm{Gyr}$. This limit was found to be $\mathrm{M}_{\mathrm{HBL}}$ = $0.066\ \mathrm{M}_\odot$ for \texttt{HMHA} (Figure~\ref{fig:prediction_ML}). For comparison, the HBL for the nominal population is at a marginally higher value of $\mathrm{M}_{\mathrm{HBL}}$ = $0.069\ \mathrm{M}_\odot$. A higher HBL mass is expected for stars with lower metallicity as the corresponding reduction in atmospheric opacity results in faster cooling and requires a higher rate of nuclear burning (higher core temperature) to sustain thermal equilibrium. We note that the hydrogen-burning limits calculated here are lower than most literature estimates (e.g. \citealt{BD_mass_limit}) due to the increased helium mass fraction that stimulates faster hydrogen fusion in the core, allowing stars of lower masses to establish energy equilibrium.

\subsection{Unresolved binary systems}

A fraction of brown dwarfs in \omegacen{} may be multiple systems which will appear brighter due to the superposition of fluxes from individual components. Existing constraints from the luminosity function \citep{binary_fraction_1} and the radial velocity distribution \citep{binary_fraction_2} suggest that \omegacen{} has an unusually low binary fraction of at most a few percent among hydrogen-burning members which is likely to be lower yet for the substellar population of the cluster \citep{2007prpl.conf..427B,2018MNRAS.479.2702F}. We may therefore safely ignore the effect of triple and higher-order systems that are far less likely to form than binary systems \citep{triple_stars}.

The effect of unresolved binary systems is determined by the binary fraction of the cluster as well as the distribution of the component mass ratio, $q=M_s/M_p$ where $M_s$ and $M_p$ are the masses of the secondary and primary components respectively and $q\leq1$. To quantify the effect, we carried out numerical simulations where a number of randomly chosen objects in the \textit{JWST} predicted dataset received secondary components with masses drawn according to the commonly used \citep{pairing_algorithms} power law distribution of mass ratios, $P(q)\propto q^\beta$. In each case, the probability distribution was trimmed at the minimum value of $q$ that ensures the secondary mass remains within the mass range of the best-fit isochrone \texttt{HMHA}. We considered a range of $\beta$ values from $\beta=-0.5$ calculated by \citet{low_q} for a variety of star forming regions, to $\beta=4$ used by \citet{2006ApJS..166..585B} for field brown dwarfs. We found the companion mass distribution at the lowest value of $q$ to closely resemble that obtained through random pairing of cluster members for our IMF. On the other hand, the highest considered value of $\beta$ emphasizes preference for components with similar masses corresponding to the so-called ``twin peaks'' effect \citep{twin_peak,pairing_algorithms}.

We found that the width of the stellar/substellar gap shown in Fig.~\ref{fig:prediction_LF} is not noticeably affected by binary systems for binary fractions under $0.5$ due to the smooth rise in brown dwarf number density with magnitude. The average brightness of modelled brown dwarfs increased by $\sim0.1\ \mathrm{mag}$ for the case of $\beta=4$ and a binary fraction of $0.2$. For the more realistic binary fraction of $0.05$, the magnitude difference did not exceed $0.03\ \mathrm{mag}$ for all considered values of $\beta$, falling well within the expected uncertainty of future $JWST$ measurements. We therefore conclude that magnitude predictions for brown dwarfs in \omegacen{} presented in this work are not noticeably affected by any realistic population of unresolved multiple star systems in the cluster.

\section{Conclusion} \label{sec:6_conclusion}
In this study, we calculated a new set of theoretical isochrones, mass-luminosity relations, and colour-magnitude diagrams for the helium-rich members of the globular cluster \omegacen{}. Our predictions provide a theoretical expectation for the first observations of brown dwarfs in globular clusters anticipated with \textit{JWST}. At present, globular cluster photometry extends below the faint end of the Main Sequence, but not deep enough to robustly sample the brown dwarf population. The predictions presented in this paper are adjusted for the metallicity and enhancements of individual elements in \omegacen{}. The necessary parameters were determined by starting with a set of abundances derived from literature spectroscopy of bright members and iteratively perturbing them until the best correspondence of the synthetic colour-magnitude diagram with the existing Main Sequence \textit{HST} photometry was achieved. Our main findings are summarized below:

\begin{itemize}
    \item In agreement with qualitative expectations, our predictions show that the Main Sequence is followed by a large stellar/substellar gap in the colour-magnitude space populated by a small number of objects. The specific size of the gap depends on the age of the cluster and the evolutionary rate of brown dwarfs, the latter of which depends on the helium mass fraction and metal abundances.

    \item The modal trend in the colour-magnitude diagram of \omegacen{} cannot be reproduced with solar or scaled solar chemical abundances as evidenced by the dependence of compatibility likelihood on enhancements of individual elements shown in Figure~\ref{fig:reddening}. For this reason, our analysis required new evolutionary interior and atmosphere models.

    \item The best-fit abundances calculated in this study are summarized in Tables~\ref{table:fixed_properties} and \ref{table:table2} corresponding to the \texttt{HMHA} population. We found that the helium-rich members are most consistent with the metal-rich end ($[\mathrm{M}/\mathrm{H}]\approx-1.4$) of the metallicity distribution in \omegacen{} in agreement with the hypothesis of \citet{bedin_bifurcation}.

    \item The best-fit isochrone, \texttt{HMHA}, is based on the distinct modal peaks of the $[\mathrm{C}/\mathrm{M}]$ and $[\mathrm{N}/\mathrm{M}]$ distributions inferred from spectroscopy of bright members in \citet{abundances}. The positions of the peaks within their distributions are consistent with the second generation of stars discussed in \citet{abundances} that is most resembling of the blue helium-rich sequence in the cluster.

    \item On the other hand, the broad $[\mathrm{O}/\mathrm{M}]$ distribution in \citet{abundances} lacked a well-defined peak. Figure~\ref{fig:iso_secondary} demonstrates that the oxygen abundance cannot be reliably constrained by our method as the optical colour-magnitude diagram of the cluster does not change significantly while $[\mathrm{O}/\mathrm{M}]$ is varied within the limits of its distribution. However, we established that the CMD depends strongly on the abundance of $\alpha$ elements. The two best-fitting secondary populations, \texttt{HMHA} and \texttt{HMMA}, both require considerable $\alpha$-enhancement with specific values of $[\alpha/\mathrm{M}]=0.6$ and $[\alpha/\mathrm{M}]=0.4$.

    \item The \textit{HST}-observed luminosity distribution of the cluster can be reproduced within uncertainties by a broken power law IMF and two populations with solar and enhanced helium mass fractions, with the latter containing fewer than $45\%$ of the members, in agreement with measurements in \citet{discrete_populations_experimental} away from the center of the cluster.

    \item We calculated the hydrogen-burning limit for the helium-rich members of \omegacen{} as $0.066\ \mathrm{M}_\odot$. This value falls below the literature predictions for a solar helium mass fraction ($\sim0.07\ \mathrm{M}_\odot$ at solar metallicity \citealt{H_limit_nominal}) 
    as larger helium mass fraction increases the core mean molecular weight, allowing faster nuclear burning and, hence, energy equilibrium in objects of lower mass.

    \item We predict that the brightest brown dwarfs in \omegacen{} will have magnitude $28$ in \textit{JWST} NIRCam \texttt{F322W2} (Figure~\ref{fig:prediction_LF}). Within our modelling range, the density of brown dwarfs appears to reach its maximum around magnitude $30$, where the brown dwarf count per magnitude is comparable to the star count per magnitude around the peak of the Main Sequence within a factor of two. 
    Based on our exposure calculations for \textit{JWST}, we predict that the brown dwarf peak is just detectable with a $1\ \mathrm{hr}$ exposure, while signal-to-noise ratios between $5$ and $10$ can be attained for the brightest brown dwarfs for the same exposure time.
\end{itemize}

The analysis in this study is based on a new set of evolutionary models and model atmospheres, which was necessitated by the significant departures of \omegacen{} abundances from the scaled solar standard that is assumed in most publicly available grids. 
Our grid reaches $T_\mathrm{eff}\sim 1\ \mathrm{kK}$ which is just sufficient to model the reappearance of brown dwarfs after the stellar/substellar gap in globular clusters. Extending the grid to even lower temperatures is currently not feasible with our present setup due to incomplete molecular opacities and associated convergence issues, requiring a future follow-up study with an improved modelling framework. 

The analysis presented here relies on the assumption that a single best-fit isochrone is sufficient to describe the average trend of \omegacen{} members in the colour-magnitude space. A more complete study must model the population with multiple simultaneous isochrones capturing the chemical complexity of the cluster that may host as many as $15$ distinct populations \citep{populations_2}. In fact, the bifurcation of the optical Main Sequence at the high temperature end into the helium-enriched and solar helium sequences is visually apparent in Figure~\ref{fig:iso_vs_data}, suggesting that the approximation is invalid in that temperature regime, which may explain the mismatch in the Main Sequence turn-off points between our best-fit prediction and the infrared dataset in the lower panel of Figure~\ref{fig:iso_vs_data}. At lower temperatures, the sequences appear more blended due to intrinsic scatter as well as increasing experimental uncertainties. However, the separation between the isochrones of different populations may be similar to or more prominent than differences around the turn-off point \citep{populations}. Other globular clusters, such as NGC~6752, also show highly distinct populations in the near infrared CMDs \citep{2019MNRAS.484.4046M}.

In this study, mixing in additional isochrones from public grids allowed us to construct a model luminosity function that approximated its observed counterpart reasonably well; however, future studies will need to produce a more extensive grid of both evolutionary and atmosphere models to capture the multiple cluster populations present. 

The current scarcity of known metal-poor brown dwarfs necessitates over-reliance on theoretical models of complex low-temperature physics that remain largely unconfirmed. The predictions drawn in this paper will be directly comparable to new globular cluster photometry expected over the next few years from both \textit{JWST} and other next generation facilities under construction. The observed size of the stellar/substellar gap as well as positions and densities of metal-poor brown dwarfs in the colour-magnitude space will then provide direct input into state-of-the-art stellar models, offering a potential to improve our understanding of molecular opacities, clouds and other low-temperature phenomena in the atmospheres of the lowest-mass stars and brown dwarfs.

\acknowledgements
This work made use of the SIMBAD database, operated at CDS, Strasbourg, France. XSEDE is supported by NSF grant ACI-1548562. Roman Gerasimov and Adam Burgasser acknowledge support from HST GO-15096.
Michele Scalco and Luigi Rolly Bedin acknowledge financial support by MIUR under PRIN program $\#$2017Z2HSMF. Maurizio Salaris acknowledges support from STFC Consolidated Grant ST/V00087X/1.

\software{
\texttt{Astropy} \citep{astropy_1,astropy_2},  
\texttt{Matplotlib} \citep{matplotlib},
\texttt{NumPy} \citep{numpy}, 
\texttt{SciPy} \citep{scipy}
}

\facilities{{\em HST} (\textit{ACS}, \textit{WFC3})}

\clearpage

\appendix

\section{Evolutionary configuration} \label{sec:8_mesa}
Table~\ref{tab:mesa} lists all \texttt{MESA v15140} settings employed in this study that differ from their default values. The initial settings were adopted from \citet{MIST}. The boundary condition tables for atmosphere-interior coupling were then replaced with the tables calculated in this study as detailed in Section~\ref{sec:3_isochrones}. Since the setup in \citet{MIST} is based on the older version of \texttt{MESA} (\texttt{v7503}), some of the settings were replaced with their modern equivalents. Finally, all parameters that have insignificantly small effect on the range of stellar masses considered in this study (e.g. nuclear reaction networks) were restored to \texttt{MESA} defaults.

\begin{deluxetable}{p{0.26\linewidth}p{0.20\linewidth}p{0.54\linewidth}}[h]
\tablenum{4}
\tablecaption{Configuration options chosen in \texttt{MESA} models calculated in this study\label{tab:mesa}}
\tablewidth{\textwidth}
\tablehead{
\colhead{Parameter} & \colhead{Value} & \colhead{Explanation}
}
\startdata
\texttt{Zbase} & Same as \texttt{initial\_z} & Nominal metallicity for opacity calculations \\
\texttt{kap\_file\_prefix} \newline \texttt{kappa\_lowT\_prefix} \newline \texttt{kappa\_CO\_prefix} & \texttt{a09} \newline \texttt{lowT\_fa05\_a09p} \newline \texttt{a09\_co} &  Opacity tables pre-computed for the solar abundances in \citet{asplund} which match the abundances adopted in this study the closest. Also following \citet{MIST} \\
\texttt{create\_pre\_main\_sequence\_model} & True & Begin evolution at the PMS, following \citep{MIST} \\
\texttt{pre\_ms\_T\_C} & $5\times10^5\ \mathrm{K}$ & Initial central temperature for the PMS, following \citep{MIST} \\
\texttt{atm\_option} & \texttt{T\_tau} for the first 100 steps and \texttt{table} after that & Boundary conditions for the atmosphere-interior coupling. Use grey atmosphere temperature relation initially, following \citep{MIST}, then switch to custom atmosphere tables \\
\texttt{atm\_table} & \texttt{tau\_100} & Use pre-tabulated atmosphere-interior coupling boundary conditions at the optical depth of $\tau=100$ \\
\texttt{initial\_zfracs} & \texttt{0} & Use custom initial abundances of elements \\
\texttt{initial\_z} & Metal mass fraction corresponding to the population of interest & $[\mathrm{M}/\mathrm{H}]$ is converted to metal mass fraction using the abundances in Tables \ref{table:fixed_properties} and \ref{table:table2} as well as solar baseline abundances in Table~\ref{tab:solar} \\
\texttt{initial\_y} & $0.4$ & Enhanced helium mass fraction, $\mathrm{Y}=0.4$, considered in this study \\
\texttt{z\_fraction\_*} & Abundances of all elements corresponding to the population of interest & Enhancements in Tables \ref{table:fixed_properties} and \ref{table:table2} as well as solar baseline abundances in Table~\ref{tab:solar} \\
\texttt{initial\_mass} & Range from $\sim0.03\ \mathrm{M}_\odot$ to $\sim0.5\ \mathrm{M}_\odot$ & Evolutionary models are calculated from the lowest mass covered by the atmosphere grid to the upper limit of $\sim0.5\ \mathrm{M}_\odot$ where the atmosphere-interior coupling scheme can no longer be used \\
\texttt{max\_age} & $13.5\ \mathrm{Gyr}$ & Terminate evolution at $13.5\ \mathrm{Gyr}$ for all stars as the maximum expected age of cluster members \\
\texttt{mixing\_length\_alpha} & $1.82$ scale heights & Convective mixing length determined by solar calibration in \citet{MIST} \\
\texttt{do\_element\_diffusion} & True & Carry out element diffusion \\
\texttt{diffusion\_dt\_limit} & $3.15\times10^7\ \mathrm{s}$ but disabled in fully convective stars & Minimum time step required by \texttt{MESA} to calculate element diffusion. The default value, $3.15\times10^7\ \mathrm{s}$, is changed to a much larger number, $3.15\times10^{16}\ \mathrm{s}$, once the mass of the convective core is within $0.01\ \mathrm{M}_\odot$ of the mass of the star to suppress diffusion in fully convective objects due to poor convergence \\
\enddata
\end{deluxetable}

\section{Solar abundances} \label{sec:7_appendix}
In this appendix, we list the solar element abundances adopted in this study for both atmosphere and evolutionary models (Table~\ref{tab:solar}). Solar abundances are presented as logarithmic ($\mathrm{dex}$) number densities compared to hydrogen whose abundance is set to $12.00\ \mathrm{dex}$ exactly. All elements omitted in the table were not included in the modeling.
The abundances listed here correspond to hydrogen, helium, and metal mass fractions of $\mathrm{X}=0.714$, $\mathrm{Y}=0.271$ and $\mathrm{Z}=0.015$ respectively.

\startlongtable
\begin{deluxetable}{ccccc|ccccc}
\tablenum{5}
\tablecaption{Solar abundances adopted in this study\label{tab:solar}}
\tablewidth{\textwidth}
\tablehead{
\colhead{Symbol} & \colhead{Element} & \colhead{Abundance} & \colhead{Error} & \colhead{Reference} & \colhead{Symbol} & \colhead{Element} & \colhead{Abundance} & \colhead{Error} & \colhead{Reference}
}
\startdata
$\mathrm{H}$ & Hydrogen & $12.00$ & -- & (1) & $\mathrm{Ru}$ & Ruthenium & $1.75$ & 0.08 & (3)  \\
$\mathrm{He}$ & Helium & $10.98$ & 0.01 & (2) & $\mathrm{Rh}$ & Rhodium & $1.06$ & 0.04 & (4)  \\
$\mathrm{Li}$ & Lithium & $3.26$ & 0.05 & (4) & $\mathrm{Pd}$ & Palladium & $1.65$ & 0.02 & (4)  \\
$\mathrm{Be}$ & Beryllium & $1.38$ & 0.09 & (3) & $\mathrm{Ag}$ & Silver & $1.20$ & 0.02 & (4)  \\
$\mathrm{B}$ & Boron & $2.79$ & 0.04 & (4) & $\mathrm{Cd}$ & Cadmium & $1.71$ & 0.03 & (4)  \\
$\mathrm{C}$ & Carbon & $8.50$ & 0.06 & (6) & $\mathrm{In}$ & Indium & $0.76$ & 0.03 & (4)  \\
$\mathrm{N}$ & Nitrogen & $7.86$ & 0.12 & (6) & $\mathrm{Sn}$ & Tin & $2.04$ & 0.10 & (3)  \\
$\mathrm{O}$ & Oxygen & $8.76$ & 0.07 & (6) & $\mathrm{Sb}$ & Antimony & $1.01$ & 0.06 & (4)  \\
$\mathrm{F}$ & Fluorine & $4.56$ & 0.30 & (3) & $\mathrm{Te}$ & Tellurium & $2.18$ & 0.03 & (4)  \\
$\mathrm{Ne}$ & Neon & $8.02$ & 0.09 & (8) & $\mathrm{I}$ & Iodine & $1.55$ & 0.08 & (4)  \\
$\mathrm{Na}$ & Sodium & $6.24$ & 0.04 & (3) & $\mathrm{Xe}$ & Xenon & $2.24$ & 0.06 & (5)  \\
$\mathrm{Mg}$ & Magnesium & $7.60$ & 0.04 & (3) & $\mathrm{Cs}$ & Caesium & $1.08$ & 0.02 & (4)  \\
$\mathrm{Al}$ & Aluminium & $6.45$ & 0.03 & (3) & $\mathrm{Ba}$ & Barium & $2.18$ & 0.09 & (3)  \\
$\mathrm{Si}$ & Silicon & $7.51$ & 0.03 & (3) & $\mathrm{La}$ & Lanthanum & $1.10$ & 0.04 & (3)  \\
$\mathrm{P}$ & Phosphorus & $5.46$ & 0.04 & (6) & $\mathrm{Ce}$ & Cerium & $1.58$ & 0.04 & (3)  \\
$\mathrm{S}$ & Sulfur & $7.16$ & 0.05 & (6) & $\mathrm{Pr}$ & Praseodymium & $0.72$ & 0.04 & (3)  \\
$\mathrm{Cl}$ & Chlorine & $5.50$ & 0.30 & (3) & $\mathrm{Nd}$ & Neodymium & $1.42$ & 0.04 & (3)  \\
$\mathrm{Ar}$ & Argon & $6.40$ & 0.13 & (5) & $\mathrm{Sm}$ & Samarium & $0.96$ & 0.04 & (3)  \\
$\mathrm{K}$ & Potassium & $5.11$ & 0.09 & (6) & $\mathrm{Eu}$ & Europium & $0.52$ & 0.04 & (3)  \\
$\mathrm{Ca}$ & Calcium & $6.34$ & 0.04 & (3) & $\mathrm{Gd}$ & Gadolinium & $1.07$ & 0.04 & (3)  \\
$\mathrm{Sc}$ & Scandium & $3.15$ & 0.04 & (3) & $\mathrm{Tb}$ & Terbium & $0.30$ & 0.10 & (3)  \\
$\mathrm{Ti}$ & Titanium & $4.95$ & 0.05 & (3) & $\mathrm{Dy}$ & Dysprosium & $1.10$ & 0.04 & (3)  \\
$\mathrm{V}$ & Vanadium & $3.93$ & 0.08 & (3) & $\mathrm{Ho}$ & Holmium & $0.48$ & 0.11 & (3)  \\
$\mathrm{Cr}$ & Chromium & $5.64$ & 0.04 & (3) & $\mathrm{Er}$ & Erbium & $0.92$ & 0.05 & (3)  \\
$\mathrm{Mn}$ & Manganese & $5.43$ & 0.04\footnote{The uncertainty in $\mathrm{Mn}$ abundance differs between the preprint (\texttt{arXiv:0909.0948}) and published versions of \citet{asplund} by $0.01\ \mathrm{dex}$. The latter is presented here.} & (3) & $\mathrm{Tm}$ & Thulium & $0.10$ & 0.04 & (3)  \\
$\mathrm{Fe}$ & Iron & $7.52$ & 0.06 & (6) & $\mathrm{Yb}$ & Ytterbium & $0.92$ & 0.02 & (4)  \\
$\mathrm{Co}$ & Cobalt & $4.99$ & 0.07 & (3) & $\mathrm{Lu}$ & Lutetium & $0.10$ & 0.09 & (3)  \\
$\mathrm{Ni}$ & Nickel & $6.22$ & 0.04 & (3) & $\mathrm{Hf}$ & Hafnium & $0.87$ & 0.04 & (6)  \\
$\mathrm{Cu}$ & Copper & $4.19$ & 0.04 & (3) & $\mathrm{Ta}$ & Tantalum & $-0.12$ & 0.04 & (4)  \\
$\mathrm{Zn}$ & Zinc & $4.56$ & 0.05 & (3) & $\mathrm{W}$ & Tungsten & $0.65$ & 0.04 & (4)  \\
$\mathrm{Ga}$ & Gallium & $3.04$ & 0.09 & (3) & $\mathrm{Re}$ & Rhenium & $0.26$ & 0.04 & (4)  \\
$\mathrm{Ge}$ & Germanium & $3.65$ & 0.10 & (3) & $\mathrm{Os}$ & Osmium & $1.36$ & 0.19 & (6)  \\
$\mathrm{As}$ & Arsenic & $2.30$ & 0.04 & (4) & $\mathrm{Ir}$ & Iridium & $1.38$ & 0.07 & (3)  \\
$\mathrm{Se}$ & Selenium & $3.34$ & 0.03 & (4) & $\mathrm{Pt}$ & Platinum & $1.62$ & 0.03 & (4)  \\
$\mathrm{Br}$ & Bromine & $2.54$ & 0.06 & (4) & $\mathrm{Au}$ & Gold & $0.80$ & 0.04 & (4)  \\
$\mathrm{Kr}$ & Krypton & $3.25$ & 0.06 & (5) & $\mathrm{Hg}$ & Mercury & $1.17$ & 0.08 & (4)  \\
$\mathrm{Rb}$ & Rubidium & $2.36$ & 0.03 & (4) & $\mathrm{Tl}$ & Thallium & $0.77$ & 0.03 & (4)  \\
$\mathrm{Sr}$ & Strontium & $2.87$ & 0.07 & (3) & $\mathrm{Pb}$ & Lead & $2.04$ & 0.03 & (4)  \\
$\mathrm{Y}$ & Yttrium & $2.21$ & 0.05 & (3) & $\mathrm{Bi}$ & Bismuth & $0.65$ & 0.04 & (4)  \\
$\mathrm{Zr}$ & Zirconium & $2.62$ & 0.06 & (7) & $\mathrm{Th}$ & Thorium & $0.08$ & 0.03 & (6)  \\
$\mathrm{Nb}$ & Niobium & $1.46$ & 0.04 & (3) & $\mathrm{U}$ & Uranium & $-0.54$ & 0.03 & (4)  \\
$\mathrm{Mo}$ & Molybdenum & $1.88$ & 0.08 & (3) &  &  &  &  &   \\
\enddata

\tablecomments{
(1) -- Hydrogen abundance is $12.00$ by definition. 
(2) -- Helium PMS abundance calibrated to the initial helium mass fraction of $Y=0.27\pm0.01$ as estimated from an ensemble of solar models in literature calibrated to observed photospheric metallicity, luminosity and helioseismic frequencies \citep{Solar_helium,Solar_helium_2}. 
(3) -- Present day spectroscopic photospheric abundances from \citet{asplund}, Table 1.
(4) -- Meteoritic abundances from \citet{asplund}, Table 1. 
(5) -- Present day indirect photospheric abundances from \citet{asplund}, Table 1.
(6) -- Present day spectroscopic photospheric abundances from \cite{extra_abundances_2}, Table 5.
(7) -- Present day spectroscopic photospheric abundance of zirconium from \citet{extra_abundances_3}, 
(8) -- Present day spectroscopic photospheric abundance of neon inferred from a representative sample of B-type stars \citep{extra_abundances_4}.}
\end{deluxetable}

\section{Catalogue} \label{sec:7b_catalog}
We include with this publication an astro-photometric catalogue of measured sources in the {\em HST} imaged fields, and multi-band atlases for each filter. The main catalogue (filename: \texttt{Catalog}) includes right ascensions and declinations in units of decimal degrees; as well as \texttt{VEGAMAG} magnitudes in \texttt{F606W}, \texttt{F814W}, \texttt{F110W} and \texttt{F160W} before zero-pointing and differential reddening corrections. The last three columns contains flags to differentiate unsaturated and saturated stars for \texttt{F606W} and \texttt{F814W} filters and a proper motion-based flag to distinguish between field stars and cluster members.

Four additional catalogues \texttt{R-I\_vs\_I.dat}, \texttt{J-H\_vs\_H.dat}, \texttt{C\_RIH\_vs\_H.dat} and \texttt{I-H\_vs\_J.dat} contain differential reddening-corrected, zero-pointed colours and magnitudes diagrams in the 
$m_{\rm F606W}-m_{\rm F814W}$ vs $m_{\rm F814W}$, 
$m_{\rm F110W}-m_{\rm F160W}$ vs $m_{\rm F160W}$, 
($m_{\rm F606W}-m_{\rm F814W})-(m_{\rm F814W}-m_{\rm F160W}$) vs $m_{\rm F160W}$, and 
$m_{\rm F814W}-m_{\rm F160W}$ vs $m_{\rm F110W}$ observational planes. All four files have the same number of entries and ordering as the main catalogue with one-to-one correspondence.

Finally, for each filter we provide two additional files containing the estimated photometric errors (\texttt{F606W\_err.dat}, \texttt{F814W\_err.dat}, \texttt{F110W\_err.dat} and \texttt{F160W\_err.dat}) and completeness (\texttt{F606W\_comp.dat}, \texttt{F814W\_comp.dat}, \texttt{F110W\_comp.dat} and \texttt{F160W\_comp.dat}) computed in each half-magnitude bin.

We also release with this publication atlases of the imaged field in each of the four filters.
These atlases consist of stacked images produced with two sampling versions: one atlas sampled at the nominal pixel resolution and one atlas sampled at 2$\times$-supersampled pixel resolution. The stacked images adhere to standard \textit{FITS} format and contain headers with astrometric \textit{WCS} (\textit{World Coordinate System}) solutions tied to \textit{Gaia} Early Data Release 3 astrometry \citep{eDR3}. 
We provide a single stacked view for each of \texttt{F606W} and \texttt{F814W} fields, 
and two stacked views for each of \texttt{F110W} and \texttt{F160W} fields separated into 
short and long exposure images.

The catalogues and atlases are included with this publication as supplementary electronic material and are available online\footnote{\href{
https://web.oapd.inaf.it/bedin/files/PAPERs_eMATERIALs/wCen_HST_LargeProgram/P05/}{https://web.oapd.inaf.it/bedin/files/PAPERs\_\\eMATERIALs/wCen\_HST\_LargeProgram/P05/}}.

\bibliography{references}{}
\bibliographystyle{aasjournal}

\end{document}